\documentclass[journal]{IEEEtran}
\usepackage{amsmath,amsfonts,bm}
\usepackage{mathrsfs}
\usepackage{newtxtext}
\usepackage{array}
\usepackage{listings}
\usepackage{adjustbox}
\usepackage[group-digits=true,group-minimum-digits=4]{siunitx}
\usepackage[caption=false,font=normalsize,labelfont=sf,textfont=sf]{subfig}
\usepackage{tikz,pgfplots}
\usepackage{textcomp}
\usepackage{stfloats}
\usepackage{url}
\usepackage{verbatim}
\usepackage{graphicx}
\usepackage{acronym}
\usepackage{caption}
\usepackage{float}
\usepackage{comment}
\usepackage{soul}
\usepackage{color}
\usepackage{mathtools}
\usepackage{graphicx}
\usepackage{balance}
\usepackage{multicol}
\usepackage{longtable}
\usepackage{amssymb}
\usepackage{booktabs, makecell, tabularx}
\usepackage[pagewise, switch]{lineno}
\usepackage{longtable}
\usepackage[section]{placeins}
\usepackage{nomencl}
\usepackage{footmisc}
\usepackage{cite}
\usepackage{xcolor}
\usepackage{caption}
\usepackage{setspace}
\usepackage{amsthm}
\usepackage{academicons}
\usepackage[hidelinks]{hyperref}
\usepackage[acronym]{glossaries}
\usepackage{mathtools, cuted}
\usepackage{lipsum}
\usepackage{balance}
\usepackage[font={footnotesize},labelfont={footnotesize}]{caption}
\usepackage{xpatch}
\usepackage{lineno}

\graphicspath{{./Pics}}

\acrodef{5g}[5G]{fifth-generation}
\acrodef{6g}[6G]{sixth-generation}
\acrodef{ao}[AO]{adaptive optics}
\acrodef{aoa}[AOA]{angle-of-arrival}
\acrodef{aoc}[AOC]{adaptive optics correction}
\acrodef{aoi}[AoI]{age of information}
\acrodef{ap}[AP]{access point}
\acrodef{aps}[APS]{active/passive switching}
\acrodef{auv}[AUV]{autonomous underwater vehicle}
\acrodef{awgn}[AWGN]{additive white Gaussian noise}
\acrodef{bb84}[BB84]{Bennett-Brassard 1984}
\acrodef{bb92}[BB92]{Bennett-Brassard 1992}
\acrodef{bep}[BEP]{bit-error probability}
\acrodef{ber}[BER]{bit-error-rate}
\acrodef{bpsk}[BPSK]{binary phase shift keying}
\acrodef{bs}[BS]{base station}
\acrodef{bsc}[BSC]{binary symmetric channel}
\acrodef{cdf}[CDF] {cumulative distribution function}
\acrodef{cflos}[CFLOS] {cloud-free line-of-sight}
\acrodef{cnc}[CNC] {cloud droplet number concentration}
\acrodef{csi}[CSI]{channel state information}
\acrodef{dl}[DL]{downlink}
\acrodef{dpsk}[DPSK]{differential phase-shift-keying}
\acrodef{ealb}[EALB]{ergodic achievable-rate lower bound}
\acrodef{egc}[EGC]{equal gain combining}
\acrodef{eh}[EH]{energy harvesting}
\acrodef{ee}[EE]{energy efficiency}
\acrodef{em}[EM]{electromagnetic}
\acrodef{es}[ES]{element splitting}
\acrodef{ess}[ES]{energy splitting}
\acrodef{et}[ET]{energy transfer}
\acrodef{eu}[EU]{energy user}
\acrodef{ew}[EW]{Exponentiated Weibull}
\acrodef{fiwi}[FiWi]{fiber-wireless}
\acrodef{fov}[FoV]{field-of-view}
\acrodef{fpa}[FPA]{fixed power allocation} 
\acrodef{fpga}[FPGA]{field-programmable gate array}
\acrodef{fso}[FSO]{free space optical}
\acrodef{gg}[GG]{Gamma-Gamma}
\acrodef{gs}[GS]{ground stations}
\acrodef{haps}[HAPS]{high altitude platform station}
\acrodef{har}[HaR]{harvest-and-reflect}
\acrodef{hris}[HRIS]{hybrid RIS}
\acrodef{hitran}[HITRAN]{HIgh resolution TRANsmission}
\acrodef{hv}[HV]{Hufnagel-Valley}
\acrodef{im/dd}[IM/DD]{intensity modulation direct detection}
\acrodef{iot}[IoT]{internet of things}
\acrodef{iout}[IoUT]{internet of underwater things}
\acrodef{irs}[IRS]{intelligent reflecting surface}
\acrodef{it}[IT]{information transfer}
\acrodef{iu}[IU]{information user}
\acrodef{isac}[ISAC]{integrated sensing and communication}
\acrodef{ld}[LD]{laser diode}
\acrodef{ldpc}[LDPC]{low density parity check}
\acrodef{led}[LED]{light-emitting diode}
\acrodef{lifi}[LiFi]{light fidelity}
\acrodef{los}[LoS]{line-of-sight}
\acrodef{lowtran}[LoWTRAN]{LOW resolution TRANsmission}
\acrodef{lwc}[LWC]{liquid water concentration}
\acrodef{mac}[MAC]{media access control}
\acrodef{map}[MAP]{maximum a posteriori}
\acrodef{mc}[MC]{Monte Carlo}
\acrodef{mcf}[MCF]{mutual coherence function}
\acrodef{mgf}[MGF]{moment generating function}
\acrodef{mimo}[MIMO]{multiple-input multiple-output}
\acrodef{miso}[MISO]{multiple-input single-output}
\acrodef{mmwave}[mmWave]{milimeter-wave}
\acrodef{modtran}[MODTRAN]{MODerate resolution TRANsmission}
\acrodef{mppm}[MPPM]{M-ary pulse position}
\acrodef{mqam}[mQAM]{multiple quadrature amplitude modulation}
\acrodef{mrc}[MRC]{maximum ratio combining}
\acrodef{ms}[MS]{Mode-Switching}
\acrodef{nccs}[NCCS]{non-central chi-square}
\acrodef{nlos}[NLoS]{non-line-of-sight}
\acrodef{noma}[NOMA]{non-orthogonal multiple access}
\acrodef{oam}[OAM]{orbital angular momentum}
\acrodef{occ}[OCC]{optical camera communication}
\acrodef{ofdm}[OFDM]{orthogonal frequency division multiplexing}
\acrodef{oma}[OMA]{orthogonal multiple access}
\acrodef{ook}[OOK]{on-off keying}
\acrodef{op}[OP]{outage probability}
\acrodef{otops}[OTOPS] {Oceanic Turbulence Optical Power Spectrum}
\acrodef{owc}[OWC]{optical wireless communication}
\acrodef{pam}[PAM]{pulse amplitude modulation}
\acrodef{papr}[PAPR]{peak-to-average power ratio}
\acrodef{pe}[PE]{pointing error}
\acrodef{pd}[PD]{photo-detector}
\acrodef{pdf}[PDF] {probability density function}
\acrodef{pep}[PEP]{pairwise error probability}
\acrodef{pin}[PIN]{positive-intrinsic-negative}
\acrodef{ppm}[PPM]{pulse position modulation}
\acrodef{ps}[PS]{power splitting}
\acrodef{psk}[PSK]{phase shift keying}
\acrodef{qam}[QAM]{quadrature amplitude modulation}
\acrodef{qber}[QBER]{quantum bit error rate}
\acrodef{qc}[QC]{quantum cryptography}
\acrodef{qkd}[QKD]{quantum key distribution}
\acrodef{qos}[QoS]{quality of service}
\acrodef{re}[RE]{reflecting element}
\acrodef{r-a}[RaA]{Reflect-and-Amplify}
\acrodef{rf}[RF]{radio frequency}
\acrodef{ri}[RI]{refractive-index}
\acrodef{ris}[RIS]{reconfigurable intelligent surface}
\acrodef{rms}[RMS]{root mean square}
\acrodef{rsma}[RSMA]{Rate-Splitting Multiple-Access}
\acrodef{rov}[ROV]{remotely operated vehicle}
\acrodef{ru}[RU]{reflection user}
\acrodef{rv}[RV]{random variable}
\acrodef{selc}[SelC]{selection combining}
\acrodef{si}[SI]{scintillation index}
\acrodef{sic}[SIC]{successive interference cancellation}
\acrodef{sim}[SIM]{subcarrier intensity modulation}
\acrodef{simo}[SIMO]{single-input multiple-output}
\acrodef{siso}[SISO]{single-input single-output}
\acrodef{sinr}[SINR]{signal-to-interference-plus-noise-ratio}
\acrodef{starris}[STAR-RIS]{simultaneous transmission and reflection reconfigurable intelligent surface}
\acrodef{skr}[SKR]{secret key rate}
\acrodef{snr}[SNR]{signal-to-noise-ratio}
\acrodef{star}[STAR]{simultaneous transmission and reflection}
\acrodef{swipt}[SWIPT]{simultaneous wireless information and power transfer}
\acrodef{ts}[TS]{time switching}
\acrodef{ttc}[TTC]{tip/tilt correction}
\acrodef{tu}[TU]{transmission user}
\acrodef{tvm}[TVM]{Time-Switching with Variable Mode-Splitting Mechanism}
\acrodef{uav}[UAV]{unmanned aerial vehicle}
\acrodef{urllc}[URLLC]{ultra-reliability low-latency communications}
\acrodef{ue}[UE]{user equipment}
\acrodef{ul}[UL]{uplink}
\acrodef{uowc}[UOWC]{underwater optical wireless communication}
\acrodef{usn}[USN]{underwater sensor network}
\acrodef{vlc}[VLC]{visible light communication}
\acrodef{wsf}[WSF]{wave structure function}
\acrodef{wsn}[WSN]{wireless sensor network}
\acrodef{wifi}[WiFi]{wireless fidelity}
\acrodef{ts}[TS]{Time-Switching}
\acrodef{ps}[PS]{Power-Splitting}

\DeclareFontFamily{U}{boondoxuprscr}{\skewchar \font =45}
\DeclareFontShape{U}{boondoxuprscr}{m}{n}{
	<-> BOONDOXUprScr-Regular}{}
\DeclareFontShape{U}{boondoxuprscr}{b}{n}{
	<-> BOONDOXUprScr-Bold}{}

\DeclareMathAlphabet{\mathboondoxo}{U}{boondoxuprscr}{m}{n}
\SetMathAlphabet{\mathboondoxo}{bold}{U}{boondoxuprscr}{b}{n}

\begin{document}
	\title{Adaptive-Optics-Aware Uplink and Downlink Analysis of Ground-to-HAPS FSO Links}
	\author{Yalçın~Ata,~\IEEEmembership{Senior Member,~IEEE},~Rudolf~Saathof,~and~Muhsin~Caner~Gökçe
		\thanks{(\textit{Corresponding author: Muhsin Caner Gökçe})}
		\thanks{Y. Ata is with the Department of Electrical and Electronics Engineering, OSTIM Technical University, OSTIM, 06374 Yenimahalle, Ankara, Turkey. Y. Ata is also with SCOPT Inc., Boston, MA, US. (e-mail:\href{mailto: ylcnata@gmail.com}{ylcnata@gmail.com})}
		\thanks{R. Saathof is with the Department of Space Engineering, Delft University of Technology, 2629 HS Delft, The Netherlands (e-mail:\href{mailto:R.Saathof@tudelft.nl}{R.Saathof@tudelft.nl}).}
		\thanks{M. C. G\"{o}k\c{c}e is with the Department of Space Engineering, Delft University of Technology, 2629 HS Delft, The Netherlands, and also with the Department of Electrical-Electronics Engineering, TED University, 06420 \c{C}ankaya, Ankara, Turkey (e-mail:\href{mailto:M.C.Gokce@tudelft.nl}{M.C.Gokce@tudelft.nl}).}
	}
	
	
	\maketitle
	
	\begin{abstract} 
		 In this paper, we present an adaptive optics (AO)-aware analytical framework for ground-to- high-altitude platform station (HAPS) free-space optical (FSO) links considering both uplink (UL) and downlink (DL) transmission. The proposed model combines atmospheric attenuation, Exponentiated Weibull (EW) turbulence, Beckmann-type pointing error (PE), and field-of-view (FoV)-limited angle-of-arrival (AOA) loss. Adaptive optics correction (AOC) is incorporated through the number of corrected modes, allowing the impact of partial and stronger wavefront compensation to be directly reflected in the channel statistics. Analytical and numerically tractable expressions are derived for the outage probability (OP), average bit-error rate (BER), and an ergodic achievable-rate lower bound (EALB) optical channel. High signal-to-noise ratio (SNR) approximations are also developed, showing that AOA/FoV blocking introduces irreducible OP and BER floors, while the decaying nonblocked component is governed by the small-channel-gain behavior of the turbulence and PE distributions. Numerical results demonstrate that the DL generally outperforms the UL because of the direction-dependent interaction with near-ground turbulence. The results also show that increasing the number of AOC modes improves OP, BER, and the EALB, although the improvement gradually saturates. The effects of average SNR, HAPS altitude, zenith angle, beam waist, boresight error, and receiver FoV are further investigated to provide practical design insight for AO-assisted ground-to-HAPS FSO systems.
	\end{abstract}
	
	\begin{IEEEkeywords}
		FSO, optical wireless communication, HAPS, adaptive optics
	\end{IEEEkeywords}
	
	\IEEEpeerreviewmaketitle
	
	\section{Introduction}
	\acused{fso}\IEEEPARstart{F}{ree-space} optical (\acs{fso}) communication offers large bandwidth, narrow beam divergence, license-free operation, high spatial reuse, and immunity to electromagnetic interference, making it attractive for beyond-5G/6G backhaul and integrated terrestrial, aerial, and space networks \cite{alimi2024revolutionizing,khalighi2014survey,kaushal2017free}. In this architecture, \acp{haps} form an intermediate layer between terrestrial infrastructure and satellites, providing wide-area coverage with shorter propagation distance, lower latency, and greater deployment flexibility than satellite systems \cite{ata2022haps}.
	
	Ground-to-\ac{haps} optical links nevertheless traverse the turbulence-dominated lower atmosphere and are impaired by absorption and scattering, turbulence-induced scintillation, beam spreading and wander, platform-induced \ac{pe}, and \ac{aoa}/\ac{fov} loss \cite{Andrews2005LaserBeam,ata2022performance,zedini2024improving}. Related aerial and vertical-link studies illustrate the importance of modeling these effects jointly. Sliti \emph{et al.} combined Beer--Lambert attenuation, scintillation, platform jitter, and finite \ac{fov} in a hybrid \ac{fso}/visible-light UAV relay \cite{sliti2026line}, while Garai \emph{et al.} evaluated ground-to-satellite modulation formats under cloud attenuation and \ac{gg} turbulence \cite{garai2025ground}. These studies motivate a composite treatment of atmospheric and alignment impairments, but do not address how an empirically motivated post-\ac{ao} irradiance distribution affects ground-to-\ac{haps} performance.
	
	Among the available mitigation methods, \ac{ao} is attractive because it directly compensates turbulence-induced wavefront distortion \cite{alimi2024revolutionizing,stotts2024Tutorial}. A wavefront sensor estimates the phase error and a controllable optical element, typically a deformable mirror, applies an approximately conjugate correction. \Ac{ttc} removes dominant wavefront-slope components, whereas higher-order modal correction mitigates progressively finer aberrations. Previous studies established the communication benefit and limitations of this approach: Tyson quantified the \ac{ber} improvement from low-order Zernike correction \cite{tyson2002bit}; Roddier related correction order to \ac{ao} gain \cite{roddier1998maximum}; Parenti \emph{et al.} showed that low-order aberrations and beam wander can invalidate conventional uplink irradiance descriptions \cite{parenti2006modeling}; and Stotts and Andrews emphasized the joint roles of scintillation, beam wander, pointing, and tracking \cite{stotts2024Tutorial}. Their field-comparison analysis further showed that \ac{ao} is most effective for weak turbulence or short ranges, whereas its benefit may reduce to \ac{ttc} or disappear over stronger, longer near-ground paths \cite{stotts2021adaptive}. Accordingly, the present communication-level model represents the aggregate \ac{ao} effect through the number of corrected modes, residual wavefront variance, scintillation index, and the resulting irradiance-distribution parameters, rather than explicitly simulating the sensor, mirror, controller, and temporal loop.
	
	For \ac{haps}-based \ac{fso} systems, Ata and Alouini analyzed attenuation, \ac{gg} turbulence, \ac{pe}, \ac{aoa} fluctuations, and \ac{aoc} for \ac{ul}, \ac{dl}, and horizontal links \cite{ata2022haps}. Their results established the importance of direction-dependent propagation and the saturation of \ac{ao} gain, but adopted the conventional \ac{gg} turbulence model. The \ac{ew} model considered here is motivated by the full-\ac{ao} field measurements in \cite{stotts2024PDFModels}, where increasing the corrected Zernike modes narrowed and shifted the received-power distribution, and the \ac{ew} law represented the measured power-in-the-bucket and power-in-the-fiber statistics - including a left-skewed full-\ac{ao} condition - more accurately than the tested lognormal and \ac{gg} models. Those measurements used a \SI{10.02}{km} daytime near-horizontal link, a \SI{10}{cm} aperture, approximately 35 corrected modes, .
	
	In \cite{priyadarshani2023earth}, authors studied an Earth-to-\ac{haps} uplink with multiple receive apertures under correlated and uncorrelated \ac{gg} turbulence and derived the corresponding \ac{op}, average \ac{ber}, diversity order, coding gain, and optimum receiver \ac{fov}. In contrast, the present work considers both \ac{ul} and \ac{dl} propagation and asks how an \ac{ao}-parameterized \ac{ew} irradiance law changes the composite-channel performance when atmospheric attenuation, Beckmann-type \ac{pe}, and \ac{fov}-limited \ac{aoa} are included. Thus, injecting the correction order into the composite channel through the dependent \ac{ew} parameters and quantifying its consequences for both propagation directions presents a new aspect of \ac{ao} dependent analysis.
	
	\subsection{Motivation and Contribution}
	Our motivation is to develop an analytically tractable, \ac{ao}-aware ground-to-\ac{haps} framework that preserves the direction-dependent \ac{ul}/\ac{dl} physics while exposing the effect of correction order on the channel statistics and performance. The main contributions are:
	\begin{itemize}
		\item A composite \ac{ul}/\ac{dl} model jointly includes attenuation, \ac{ao}-dependent \ac{ew} turbulence, Beckmann-type \ac{pe}, and \ac{aoa}/\ac{fov} loss to capture their different interaction with near-ground turbulence.
		\item Analytical expressions are derived for the channel distribution, \ac{op}, average \ac{ber}, and \ac{ealb}.
		\item High-\ac{snr} analysis is performed to show approximations and diversity order.
		\item Numerical results are verified with \ac{mc} simulation.
	\end{itemize}
	
	Sections~\ref{sec:sys_model} and \ref{sec:chan_model} present the system and channel models. Sections~\ref{sec:perf_analysis} and \ref{sec:highsnr} derive the exact and high-\ac{snr} performance expressions, respectively; Section~\ref{sec:results_discussions} discusses the numerical results, and Section~\ref{sec:conclusion} concludes the paper.
	
	\section{System  Model}\label{sec:sys_model}
	\begin{figure}[!t]
		\centering
		\includegraphics[width=\columnwidth]{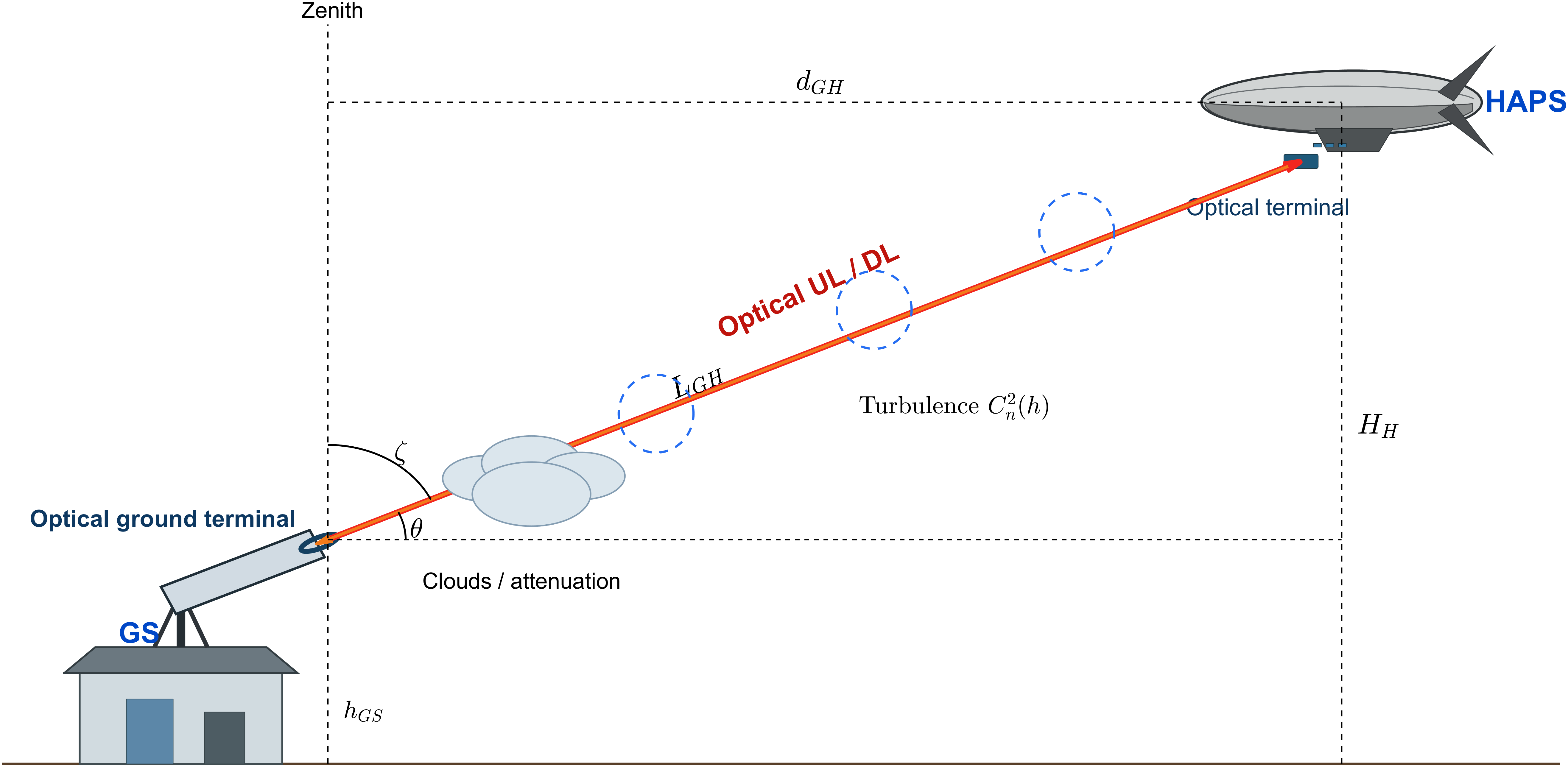}
		\caption{Receiver-side-\ac{ao}-assisted optical \ac{ul} and \ac{dl} between the ground station and the \ac{haps}.}
		\label{fig:sys_model}
	\end{figure}
	
	We consider an \ac{aoc}-assisted \ac{fso} communication system between a \ac{gs} and a \ac{haps}, as illustrated in Fig.~\ref{fig:sys_model}. The \ac{haps} operates at an altitude of $H_H$ above the ground level, typically in the range of $18$--$28$ km. The \ac{gs} terminal is located at a height $h_{GS}$ above the ground surface. The horizontal and slant separation distance between the \ac{gs} and the \ac{haps} are denoted by $d_{GH}$ and $L_{GH}= \sqrt{d_{GH}^{2} + \left(H_H-h_{GS}\right)^2}$, respectively.
	
	The elevation angle $\theta$, measured from the local horizontal plane at the \ac{gs}, is given by $\theta = \tan^{-1}[({H_H-h_{GS}})/{d_{GH}}]$,
whereas the zenith angle is $\zeta = {\pi}/{2}-\theta$.
	
    Although the same geometrical path length $L_{GH}$ is considered for both directions, the turbulence effect may differ due to the direction-dependent interaction between the optical beam and altitude-varying atmospheric turbulence. In particular, the \ac{ul} is generally more sensitive to near-ground turbulence because the beam is distorted close to the transmitter and the resulting distortion accumulates along the propagation path. In contrast, in the \ac{dl}, turbulence-induced scintillation, beam spreading, and \ac{aoa} fluctuations mainly affect the beam as it approaches the \ac{gs} receiver aperture.

    	The analysis assumes receiver-side \ac{ao} at the destination of each propagation direction: an \ac{ao}-equipped optical receiver on the \ac{haps} for the \ac{ul}, and an \ac{ao}-equipped optical ground receiver for the \ac{dl}. 
        Unless a direction is stated explicitly, the subsequent channel and performance expressions apply to both links after evaluating the direction-dependent propagation integrals and receiver-side parameters for the corresponding \ac{ul} or \ac{dl} configuration.
	
	The atmospheric turbulence strength along the propagation path is characterized by the \ac{ri} structure parameter $C_n^2$. In addition to turbulence, the received optical signal is affected by atmospheric attenuation caused by absorption, scattering, fog, clouds, or haze. Platform-induced \ac{pe} is included separately and represents displacement caused by platform motion, tracking imperfections, mechanical vibration, and alignment bias. 
	
	In practice, \ac{aoc} reduces wavefront aberrations, improves beam focusing at the receiver plane, and mitigates turbulence-induced intensity fluctuations. Therefore, the considered system model captures the main propagation impairments of \ac{gs}--\ac{haps} \ac{fso} links together with the potential performance improvement provided by \ac{aoc} assistance. 
	
	\section{Channel  Model}\label{sec:chan_model}
	
	\subsection{Attenuation}
	According to the Beer--Lambert law, attenuation $h_{al}$ caused by absorption and scattering is expressed as \cite{Vasylyev2019Satellite-mediatedLinks}
	\begin{equation}
		h_{al} = \exp\left[-\int_0^{L_{GH}} C(z) dz\right],
		\label{eq:beer_lambert_attenuation}
	\end{equation}
	where $z$ denotes the distance parameter  and 
	\begin{equation}
		C(z) = C_0(\lambda)\exp(-z/H_0\sec(\zeta))
	\end{equation}
	$C_0(\lambda)$ is the attenuation coefficient, $H_0$ a scaling parameter. 
    $C_0(\lambda)$ is expressed as a function of the wavelength $\lambda$ in nm and visibility $V$ in km as \cite{AlNaboulsi2004Fog}
	\begin{equation}
		C_0(\lambda)=\frac{3.912}{V}\left(\frac{\lambda}{550}\right)^{-q},
		\label{eq:kim_attenuation}
	\end{equation}
	where $q$ is the size-distribution coefficient and is given by \cite{Awan2009Cloud}
	\begin{equation}
		q =
		\begin{cases}
			1.6, & V > 50~\mathrm{km},\\
			1.3, & 6~\mathrm{km} < V < 50~\mathrm{km},\\
			0.16V + 0.34, & 1~\mathrm{km} < V < 6~\mathrm{km},\\
			V - 0.5, & 0.5~\mathrm{km} < V < 1~\mathrm{km},\\
			0, & V < 0.5~\mathrm{km}.
		\end{cases}
		\label{eq:kim_q}
	\end{equation} 
    Since the optical beam propagates through the troposphere before reaching the \ac{haps}, cloud layers may appear along the propagation path. Cloud tops may extend to approximately $10$--$15~\mathrm{km}$ depending on cloud type. To include the cloud effect, the visibility is related to the \ac{lwc} and the \ac{cnc} as \cite{Awan2009Cloud}
	\begin{equation}
		V = \frac{1.002}{\left(\mathrm{LWC}\times \mathrm{CNC}\right)^{0.6473}}.
		\label{eq:cloud_visibility}
	\end{equation}
	
	\subsection{Pointing Error}
	\ac{pe} is another important impairment in \ac{fso} links, especially for \ac{gs}--\ac{haps} communication, where platform motion, tracking imperfections, mechanical vibrations, and beam wander may cause a displacement between the optical beam footprint and the receiver aperture center. According to Beckmann distribution, the \ac{pdf} of the radial beam displacement $r$ is given by \cite{Simon2005Digital,BoludaRuiz2016Beckmann}
	\begin{equation}
		\begin{aligned}
			f_r(r)
			= &\frac{r}{2\pi\sigma_x\sigma_y}
			\int_{0}^{2\pi}
			\exp\Bigg[
			-\frac{\left(r\cos\varphi-\mu_x\right)^2}{2\sigma_x^2} \\
			&-\frac{\left(r\sin\varphi-\mu_y\right)^2}{2\sigma_y^2}
			\Bigg] d\varphi, \qquad r \geq 0,
		\end{aligned}
		\label{eq:beckmann_pdf}
	\end{equation}
	where $r=\sqrt{x^2+y^2}$ is the radial displacement. Here, $x$ and $y$ denote the horizontal and vertical misalignment components, respectively, and they are modeled as independent Gaussian random variables with means $\mu_x$ and $\mu_y$, and variances $\sigma_x^2$ and $\sigma_y^2$, respectively.
	
	The \ac{pe}-induced fading coefficient is expressed as \cite{Farid2007Outage}
	\begin{equation}
		h_{pl}=A_0\exp\left(-\frac{2r^2}{\omega_e^2}\right),
		\label{eq:pointing_fading}
	\end{equation}
	where $A_0$ denotes the fraction of the collected power at zero radial displacement and $\omega_e$ is the equivalent beamwidth at the receiver plane. These parameters are defined as
	\begin{equation}
		\begin{aligned}
			A_0 &= \left[\operatorname{erf}(v)\right]^2,\, 
			\omega_e = \omega_b
			\left[
			\frac{\sqrt{\pi}\operatorname{erf}(v)}
			{2v\exp(-v^2)}
			\right]^{\frac{1}{2}},\,
			v = \frac{\sqrt{\pi} r_a}{\sqrt{2}\omega_b},
		\end{aligned}
		\label{eq:pointing_parameters}
	\end{equation}
	where $\operatorname{erf}(\cdot)$ is the error function, $r_a=D_G/2$ is the receiver aperture radius, $D_G$ is the aperture diameter, and $\omega_b$ is the beam waist at the receiver plane. The \ac{pdf} of the \ac{pe}-induced fading coefficient $h_{pl}$ can be approximated as \cite{BoludaRuiz2016Beckmann}
	\begin{equation}
		f_{h_{pl}}\left(h_{pl}\right)
		=
		\frac{\phi_{\mathrm{mod}}^2}
		{\left(A_0G\right)^{\phi_{\mathrm{mod}}^2}}
		h_{pl}^{\phi_{\mathrm{mod}}^2-1},
		\quad
		0 \leq h_{pl} \leq A_0G,
		\label{eq:pointing_pdf}
	\end{equation}
	where $\phi_{\mathrm{mod}}=\omega_e/({2\sigma_{\mathrm{mod}}})$
	and the correction factor $G$ is
	\begin{equation}
		\begin{split}
			G
			= &\exp\Bigg(
			\frac{1}{\phi_{\mathrm{mod}}^2}
			-\frac{1}{2\phi_x^2}
			-\frac{1}{2\phi_y^2}
			-\frac{\mu_x^2}{2\sigma_x^2\phi_x^2} -\frac{\mu_y^2}{2\sigma_y^2\phi_y^2}
			\Bigg),
		\end{split}
		\label{eq:G_factor}
	\end{equation}
	where $\phi_x=\omega_e/(2\sigma_x)$ and $\phi_y=\omega_e/(2\sigma_y)$ represent the normalized beamwidth parameters in the horizontal and vertical directions. The modified jitter variance $\sigma_{\mathrm{mod}}^2$ is defined as
	\begin{equation}
		\sigma_{\mathrm{mod}}^2
		=
		\sqrt[3]{
			(
			3\mu_x^2\sigma_x^4
			+
			3\mu_y^2\sigma_y^4
			+
			\sigma_x^6
			+
			\sigma_y^6)
			/2
		}.
		\label{eq:sigma_mod}
	\end{equation}
	
	It is emphasized that Eq.~\eqref{eq:pointing_pdf} is the modified-Rayleigh approximation of the Beckmann \ac{pe} model developed in \cite{BoludaRuiz2016Beckmann}. The approximation simultaneously accounts for unequal horizontal and vertical jitter variances and nonzero boresight displacements through $\sigma_{\mathrm{mod}}^2$ and $G$. Its accuracy was verified in \cite{BoludaRuiz2016Beckmann} by comparing analytical \ac{op} and asymptotic-\ac{ber} results with \ac{mc} simulations.

    	Atmospheric beam wander, \ac{aoa} fluctuations, and platform-induced \ac{pe} are treated as three physically distinct mechanisms. Beam wander enters the tip/tilt-tracked scintillation model through the residual beam-wander variance $\sigma_{pe\text{-}\mathrm{TT}}^2$ in Eq.~\eqref{eq:sigmaI_TT} and its effect is consequently included in $h_{\mathrm{at}}$. 
        $h_{\mathrm{pl}}$ represents platform vibration, attitude fluctuations, tracking jitter, and deterministic alignment bias through $\mu_x$, $\mu_y$, $\sigma_x$, and $\sigma_y$.

	When the optical beam arrives at the receiver with a nonzero detection angle $\theta_d$, the collected power is further affected by the projected effective aperture. Therefore, the conditional PDF of $h_{pl}$ for a given channel state can be expressed as
	\begin{equation}
		f_{h_{pl}\mid\theta_d}\left(h_{pl}\right)
		=
		\frac{\phi_{\mathrm{mod}}^2}
		{\left(A_0G\cos\theta_d\right)^{\phi_{\mathrm{mod}}^2}}
		h_{pl}^{\phi_{\mathrm{mod}}^2-1},
		\label{eq:conditional_pointing_pdf}
	\end{equation}
	where $0 \leq h_{pl} \leq A_0G\cos\theta_d$. 
	\subsection{\ac{aoa} Fluctuations}
    Due to wavefront distortions and beam wander, the received optical beam may arrive at the receiver aperture with a random angular deviation. This angular deviation can reduce the effective collected optical power, especially when the receiver has a limited \ac{fov}. Assuming that $\theta_d$ follows a Rayleigh distribution, the PDF of $\theta_d$ is given by \cite{Safi2020GroundHAP}
	\begin{equation}
		f_{\theta_d}(\theta_d)
		=
		\frac{\theta_d}{\sigma_0^2}
		\exp\left(-\frac{\theta_d^2}{2\sigma_0^2}\right),
		\qquad \theta_d \geq 0,
		\label{eq:aoa_rayleigh_pdf}
	\end{equation}
	where $\sigma_0^2$ is the variance of the underlying zero-mean Gaussian-distributed angular deviation components.
	
	\subsection{Atmospheric Turbulence Modeling with \ac{aoc}}
	It has been shown in \cite{stotts2024PDFModels} that the \ac{ew} distribution provides an effective model for full \ac{aoc}, particularly since the corresponding received-power distribution may exhibit a left-skewed behavior under full \ac{ao} operation. Accordingly, the turbulence \ac{pdf}, whose \ac{ao} dependence is reflected through the reduced scintillation and improved coupling characteristics, can be expressed with related parameters as follows
	\begin{equation}
		\label{eq:EW_pdf}
		\begin{split}
			f_{h_{\mathrm{at}}}(h_{\mathrm{at}})
			&=
			\frac{\alpha\eta}{\beta}
			\left(\frac{h_{\mathrm{at}}}{\beta}\right)^{\alpha-1}
			\exp\!\left[-\left(\frac{h_{\mathrm{at}}}{\beta}\right)^{\alpha}\right] \\
			&\times
			\left\{1-\exp\!\left[-\left(\frac{h_{\mathrm{at}}}{\beta}\right)^{\alpha}\right]\right\}^{\eta-1},
			\qquad h\ge 0,
		\end{split}
	\end{equation}
	\begin{equation}
		\label{eq:eta}
		\eta = 3.931\left(\frac{D_{G}}{\rho_{0}}\right)^{-0.519},\quad \alpha = \left[\eta \sigma_{I,AO}^{2}(N)\right]^{-6/11},
	\end{equation}
	\begin{equation}
		\label{eq:beta}
		\beta=
		\frac{1}{\eta\Gamma\!\left(1+\frac{1}{\alpha}\right)
			g(\eta,\alpha)},
	\end{equation}
	\begin{equation}
		\label{eq:gfunc}
		g(\eta,\alpha)
		=
		\sum_{n=0}^{\infty}
		\frac{(-1)^n(n+1)^{-(1+1/\alpha)}\Gamma(\eta)}
		{n!\,\Gamma(\eta-n)},
	\end{equation}
where $\rho_0$ is the coherence length isgiven in Eq. \eqref{eq:rho0}, $N$ is the number of \ac{aoc} modes and $\Gamma(.)$ is Gamma function. It was reported that  $g(\eta,\alpha)$ converges for  ten terms \cite{barrios2012exponentiated}. 
	The tip/tilt-tracked \ac{si} is modeled as \cite{stotts2024Tutorial}
	\begin{equation}
		\label{eq:sigmaI_TT}
		\begin{split}
			&\sigma_{I}^{2}(D_G)
			=
			-1+34.29\left(\frac{\Lambda L_{GH}}{k r_{0\mathrm{rx}}^{2}}\right)^{5/6}
			\frac{\sigma_{pe\text{-}\mathrm{TT}}^{2}}{D_{G}^{2}/8} \\
			&+ \exp\!\left\{
			\frac{0.49\sigma_{B}^{2}}
			{\left[1+0.56(1+\Theta)\sigma_{B}^{12/5}\right]^{7/6}}
			+
			\frac{0.51\sigma_{B}^{2}}
			{\left(1+0.69\sigma_{B}^{12/5}\right)^{5/6}}
			\right\},
		\end{split}
	\end{equation}
	where $k=2\pi / \lambda$, \(\Lambda=2L_{GH}/(kW_{0}^2)\), and
	\(\Theta=1-L_{GH}/F_{0}\). Here, \(W_{0}\) and \(F_{0}\)
	denote the transmitter beam radius and radius of curvature,
	respectively. In Eq. \eqref{eq:sigmaI_TT}, $r_{0\mathrm{rx}}$ is the Fried's parameter and it is given for receiver-side by \cite{stotts2024Tutorial}
	\begin{equation}
		\label{eq:r0rx}
		r_{0\mathrm{rx}}
		=
		\left[
		\frac{16.71}{\lambda^2}\sec(\zeta)\int\limits_{0}^{L_{GH}}
		C_n^2(z)\left(z/L_{GH}\right)^{5/3}\,dz
		\right]^{-\frac{3}{5}}.
	\end{equation}
	
	The tip/tilt residual \ac{pe} variance in Eq. \eqref{eq:sigmaI_TT} is given by \cite{stotts2024Tutorial}
	\begin{equation}
		\label{eq:sigmape_tt}
		\sigma_{pe\text{-}\mathrm{TT}}^{2}
		=
		\left(\sqrt{\langle r_{c}^{2}\rangle}-T_{z}L_{GH}\right)^2
		\left(1-\frac{Q}{1+Q}\right)^{1/6},
	\end{equation}
	\begin{equation}
		\label{eq:Qj}
		Q=\pi^2\left(\frac{D_{t}}{2\sqrt{2}\,r_{0\mathrm{tx}}}\right)^2,
	\end{equation}
	\begin{equation}
		\label{eq:rc2}
		\langle r_{c}^{2}\rangle
		=
		0.54\left(\frac{\lambda L_{GH}}{D_{t}/\sqrt{2}}\right)^2
		\left(\frac{D_{t}/\sqrt{2}}{r_{0\mathrm{tx}}}\right)^{5/3},
	\end{equation}
	\begin{equation}
		\label{eq:Tz}
		T_{z}
		=
		0.57\left(\frac{\lambda}{D_{t}/\sqrt{2}}\right)
		\left(\frac{D_{t}/\sqrt{2}}{r_{0\mathrm{tx}}}\right)^{5/6},
	\end{equation}
	where $D_t$ is the transmitter aperture diameter, and $r_{0\mathrm{tx}}$ is the Fried's parameter for spherical wave at the transmitter \cite{stotts2024Tutorial}
	\begin{equation}
		\label{eq:r0tx}
		r_{0\mathrm{tx}}
		=
		\left[
		\frac{16.71\sec(\zeta)}{\lambda^2}\,
		\int\limits_{0}^{L_{GH}}
		C_n^2(z)\left(\dfrac{L_{GH}-z}{L_{GH}}\right)^{\frac{5}{3}}\,dz
		\right]^{-\frac{3}{5}}.
	\end{equation}
	 Rytov variance of Gaussian beam $\sigma_B^2$ in Eq.\eqref{eq:sigmaI_TT} is
	\begin{equation}
		\label{eq:sigmaB2}
		\begin{split}
			\sigma_{B}^{2}
			&= 8.70\,k^{7/6}(H_H-h_{GS})^{5/6}\sec^{11/6}(\zeta) \\
			&\quad \times \Re \Bigg\{
			\int_{h_{GS}}^{H_H} C_n^2(h)
			\Bigg[
			\xi^{5/6}
			\Big(
			\Lambda \xi + i\big(1-\bar{\Theta}\xi\big)
			\Big)^{-5/6} \\
			&\quad 
			-\Lambda^{5/6}\xi^{5/3}
			\Bigg] dh
			\Bigg\},
		\end{split}
	\end{equation}
	where $h$ is height parameter, $\bar{\Theta}=1-\Theta$, $\xi$ is the normalized distance parameter and it is given for \ac{dl} and \ac{ul} by
	\begin{equation}
		\label{eq:xi_def}
		\xi=\frac{h-h_{GS}}{H_H-h_{GS}}, \qquad
		\xi=1-\frac{h-h_{GS}}{H_H-h_{GS}}.
	\end{equation}
	In Eq. \eqref{eq:sigmaB2}, $C_n^2(h)$ is the \ac{ri} structure parameter and according to the \ac{hv} model, it is given by \cite[pp.522]{Andrews2005LaserBeam}
	\begin{equation}
		\label{eq:Cn2_HV}
		\begin{split}
			&C_n^2(h) =
			0.00594\left(\frac{w}{27}\right)^2 (10^{-5}h)^{10}
			\exp\!\left(-\frac{h}{1000}\right)  \\
			&+ 2.7\times10^{-16}\exp\!\left(-\frac{h}{1500}\right)
			+ A\exp\!\left(-\frac{h}{100}\right),
		\end{split}
	\end{equation}
	where \(w\) is the \ac{rms} wind speed in m/s, and \(A=C_n^2(0)\) is the nominal ground-level value.

	The \ac{ul} and \ac{dl} employ the same altitude-dependent $C_n^2(h)$ profile; however,  
    their physical distinction is represented by the normalized distance parameter in the direction-dependent turbulence integrals, so that the same $C_n^2(h)$ layers receive different geometrical propagation weights in the \ac{ul} and \ac{dl}.
	
	For an \ac{aoc} system with \(N\) corrected modes, the residual wavefront variance is approximated as \cite{roddier1998maximum}
	\begin{equation}
		\label{eq:sigmawf}
		\sigma_{wf}^{2}(N)
		=
		0.335\,N^{-5/6}
		\left(\frac{D_{G}}{r_{0\mathrm{rx}}}\right)^{5/3}.
	\end{equation}
	Thus, the \ac{aoc}-dependent \ac{si} can be tailored depending on the tip/tilt tracked \ac{si} as
	\begin{equation}
		\label{eq:sigmaI_AO}
		\sigma_{I,\mathrm{AO}}^{2}(N)
		=
		\frac{\sigma_{wf}^{2}(N)}{\sigma_{wf}^{2}(3)}
		\,\sigma_{I}^{2}(D_G)
		=
		\left(\frac{N}{3}\right)^{-5/6}\sigma_{I}^{2}(D_G),
	\end{equation}
	where $N=3$ corresponds to \ac{ttc}. 
      The central assumption in Eq.\eqref{eq:sigmaI_AO} is that the fractional reduction in residual wavefront variance relative to \ac{ttc} produces the same fractional reduction in the aperture-averaged \ac{si}. This assumption provides a tractable mapping from the number of corrected modes to the parameters of the \ac{ew} irradiance distribution. The model consequently represents the aggregate effect of \ac{ao} through $N$ and a scalar residual \ac{si}. 
The empirical motivation for the scaling is provided by \cite{stotts2024PDFModels}, where the residual-wavefront-variance reduction between \ac{ttc} ($N=3$) and full \ac{ao} ($N\approx35$) was applied to the \ac{si}, and the resulting \ac{ew} PDFs showed reasonable agreement with measured power-in-the-bucket and power-in-the-fiber distributions. The supporting measurements correspond to a \SI{10.02}{km} daytime near-horizontal link, a \SI{10}{cm} receiver aperture, and $D_{\mathrm{rx}}/r_0=2.1$ and $4.25$. 

	Thus, the physical \ac{ao} dependence is represented by the chain $N\rightarrow\sigma_{wf}^{2}(N)\rightarrow\sigma_{I,\mathrm{AO}}^{2}(N)\rightarrow\{\alpha(N),\beta(N)\}$. The scalar residual wavefront variance is explicitly represented by Eq.~\eqref{eq:sigmawf}; $N=3$ denotes \ac{ttc}, whereas $N>3$ represents progressively higher-order modal correction. 
	
	The coherence length \(\rho_{0}\) in Eq. \eqref{eq:eta} is given as \cite[pp. 209]{Andrews2005LaserBeam}
	\begin{equation}
		\label{eq:rho0}
		\begin{split}
			\rho_{0}
			&=
			\Bigg[
			1.46k^2\int_{0}^{L_{GH}} C_n^2(z)
			\Bigg(
			\left|\Theta+\bar{\Theta}\frac{z}{L_{GH}}\right|^{5/3}  \\
			&+0.62\,\Lambda^{11/6}\left(1-\frac{z}{L_{GH}}\right)^{5/3}
			\Bigg)\,dz
			\Bigg]^{-3/5}.
		\end{split}
	\end{equation}
	\section{Performance Analysis} \label{sec:perf_analysis}
	\subsection{Channel \ac{pdf}}
	Combining the effects of attenuation, atmospheric
	turbulence, \ac{pe}, and \ac{aoa} fluctuations, the overall channel
	coefficient for the ground-to-\ac{haps} link can be expressed as
	\begin{equation}
		\label{eq:overall_channel}
		h = h_{\mathrm{al}} h_{\mathrm{at}} h_{\mathrm{pl}} h_{\mathrm{af}} .
	\end{equation}
	Let the composite channel excluding the \ac{aoa} term be defined as $h_{\mathrm{ag}} = h_{\mathrm{al}} h_{\mathrm{at}} h_{\mathrm{pl}}$. Then, by conditioning the attenuation
	\(h_{\mathrm{al}}\), the turbulence component
	\(h_{\mathrm{at}}\), and the \ac{pe} component
	\(h_{\mathrm{pl}}\) on \(h_{\mathrm{af}}\), the conditional \ac{pdf} of
	\(h_{\mathrm{ag}}\) can be written as
	\begin{equation}
		\label{eq:fhag_cond_2}
		f_{h_{\mathrm{ag}}|h_{\mathrm{af}}}(h_{\mathrm{ag}})
		=
		\int_{\Delta_1}^{\infty}
		\frac{1}{h_{\mathrm{al}} h_{\mathrm{at}}}
		f_{h_{\mathrm{pl}}|\theta_d}
		\!\left(
		\frac{h_{\mathrm{ag}}}{h_{\mathrm{al}} h_{\mathrm{at}}}
		\right)
		f_{h_{\mathrm{at}}}(h_{\mathrm{at}})
		\, dh_{\mathrm{at}},
	\end{equation}
	where $\Delta_1=\frac{h_{\mathrm{ag}}}{A_0 G h_{\mathrm{al}} \cos{\theta_d}}$. Inserting Eqs. \eqref{eq:conditional_pointing_pdf} and \eqref{eq:EW_pdf} into Eq. \eqref{eq:fhag_cond_2}, we have
	\begin{equation}
		\label{eq:fhag_cond_3}
		\begin{split}
			f_{h_{\mathrm{ag}}|\theta_d}(h_{\mathrm{ag}})
			&=
			\frac{\phi_{\mathrm{mod}}^{2}\,h_{\mathrm{ag}}^{\phi_{\mathrm{mod}}^{2}-1}}
			{\left(h_{\mathrm{al}}A_{0}G\cos\theta_{d}\right)^{\phi_{\mathrm{mod}}^{2}}}
			\frac{\alpha\eta}{\beta^{\alpha}}\\
			&\times
			\int_{\Delta_1}^{\infty}
			h_{\mathrm{at}}^{\alpha-\phi_{\mathrm{mod}}^{2}-1}
			\exp\!\left[-\left(\frac{h_{\mathrm{at}}}{\beta}\right)^{\alpha}\right]
			\\
			&\quad \times
			\left\{1-\exp\!\left[-\left(\frac{h_{\mathrm{at}}}{\beta}\right)^{\alpha}\right]\right\}^{\eta-1}
			\, dh_{\mathrm{at}} .
		\end{split}
	\end{equation}
	Applying Newton's generalized binomial theorem
	to the last term in Eq.\eqref{eq:fhag_cond_3}  
	\begin{equation}
		\label{eq:fhag_theta_final}
		\begin{split}
			f_{h_{\mathrm{ag}}|\theta_d}(h_{\mathrm{ag}})
			&=
			\frac{\eta \phi_{\mathrm{mod}}^{2}\,
				h_{\mathrm{ag}}^{\phi_{\mathrm{mod}}^{2}-1}}
			{\left(h_{\mathrm{al}}A_{0}G\beta\cos\theta_d\right)^{\phi_{\mathrm{mod}}^{2}}}\\
			&\times
			\sum_{n=0}^{\infty}
			\frac{(-1)^n\Gamma(\eta)}
			{n!\,\Gamma(\eta-n)\,(1+n)^{\frac{\alpha-\phi_{\mathrm{mod}}^{2}}{\alpha}}}
			\\
			&\times
			\Gamma\!\left(
			\frac{\alpha-\phi_{\mathrm{mod}}^{2}}{\alpha},
			\frac{1+n}{\beta^{\alpha}}
			\Delta_1^{\alpha}
			\right),
		\end{split}
	\end{equation}
	where \(\Gamma(\cdot,\cdot)\) is the upper incomplete Gamma function and it can be expressed
with Meijer-\(G\) function as $\Gamma(a,z)=G_{1,2}^{2,0}\!\big(
	z \,|\, \begin{matrix} 1 \\ 0,\; a \end{matrix}
	\big)$ \cite[Eq.~(06.06.26.0005.01)]{Wolfram}. Then, 
	\begin{equation}
		\label{eq:fhag_theta_meijerg}
		\begin{split}
			f_{h_{\mathrm{ag}}|\theta_d}&(h_{\mathrm{ag}})
			=
			\frac{\eta \phi_{\mathrm{mod}}^{2}\,
				h_{\mathrm{ag}}^{\phi_{\mathrm{mod}}^{2}-1}}
			{\left(h_{\mathrm{al}}A_{0}G\beta\cos\theta_d\right)^{\phi_{\mathrm{mod}}^{2}}}\\
			&\times
			\sum_{n=0}^{\infty}
			\frac{(-1)^n\Gamma(\eta)}
			{n!\,\Gamma(\eta-n)\,(1+n)^{\frac{\alpha-\phi_{\mathrm{mod}}^{2}}{\alpha}}}
			\\
			&\times
			G_{1,2}^{2,0}\!\left(
			\frac{(1+n)\,h_{\mathrm{ag}}^{\alpha}}
			{\left(\beta A_{0}G h_{\mathrm{al}}\cos\theta_d\right)^{\alpha}}
			\,\middle|\,
			\begin{matrix}
				1 \\
				0,\; \dfrac{\alpha-\phi_{\mathrm{mod}}^{2}}{\alpha}
			\end{matrix}
			\right).
		\end{split}
	\end{equation}
	
	The \ac{pdf} of the overall channel state can be obtained as
	\begin{equation}
		\label{eq:fh_def}
		\begin{split}
			f_h(h)
			=&
			\int\limits_{0}^{\theta_{\mathrm{FoV}}}
			f_{h_{\mathrm{ag}}|\theta_d}(h)\,f_{\theta_d}(\theta_d)\,d\theta_d\\
            &+
			\delta(h)\int\limits_{\theta_{\mathrm{FoV}}}^{\infty}
			f_{\theta_d}(\theta_d)\,d\theta_d, 
		\end{split}
	\end{equation}
	where $\theta_{FOV}$ is the \ac{fov}.
	Substituting Eqs. \eqref{eq:aoa_rayleigh_pdf} and
	\eqref{eq:fhag_theta_meijerg} into Eq. \eqref{eq:fh_def}, we obtain
	\begin{equation}
		\label{eq:fh_final}
		\begin{split}
			&f_h(h)
			=
			\frac{\eta \phi_{\mathrm{mod}}^{2} h^{\phi_{\mathrm{mod}}^{2}-1}}
			{\left(h_{\mathrm{al}}A_0G\beta\cos{\theta_d}\right)^{\phi_{\mathrm{mod}}^{2}}}
			\sum_{n=0}^{\infty}
			\frac{(-1)^n\Gamma(\eta)}
			{n!\,\Gamma(\eta-n)\,(1+n)^{\frac{\alpha-\phi_{\mathrm{mod}}^{2}}{\alpha}}}
			\\
			& \times
			\int_{0}^{\theta_{\mathrm{FoV}}}
			G_{1,2}^{2,0}\!\left(
			\frac{(1+n)h^{\alpha}}
			{\left(\beta A_0G h_{\mathrm{al}}\cos\theta_d\right)^{\alpha}}
			\,\middle|\,
			\begin{matrix}
				1 \\
				0,\; \dfrac{\alpha-\phi_{\mathrm{mod}}^{2}}{\alpha}
			\end{matrix}
			\right)
			\\
			&\times
			\frac{\theta_d}{\sigma_0^2}
			\exp\!\left(-\frac{\theta_d^2}{2\sigma_0^2}\right)
			\,d\theta_d
			+
			\delta(h)\int_{\theta_{\mathrm{FoV}}}^{\infty}
			\frac{\theta_d}{\sigma_0^2}
			\exp\!\left(-\frac{\theta_d^2}{2\sigma_0^2}\right)
			\,d\theta_d .
		\end{split}
	\end{equation}
	To evaluate the \(\theta_d\)-dependent first integral in Eq. \eqref{eq:fh_final},
	we employ the small-angle approximation \(\cos(\theta_d)\approx 1\),
	which has been directly assessed in \cite{Zedini2025Beckmann}. Ref. \cite{Zedini2025Beckmann} compares the \ac{op} obtained with $\cos(\theta_d)\approx1$ against a numerical evaluation retaining the exact cosine term, while the corresponding curves are visually indistinguishable and the approximation introduces negligible error for practical, small receiver-\ac{fov} angles.
	Then, applying Eqs. (3.381-8) and (3.381-9) of \cite{Gradshteyn2014Table} to the first and second integrals, respectively, the channel \ac{pdf} is obtained as
	\begin{equation}
		\label{eq:fh_small_angle}
		\begin{split}
			f_h(h)
			&=
			\frac{\eta \phi_{\mathrm{mod}}^{2} h^{\phi_{\mathrm{mod}}^{2}-1}}
			{\left(h_{\mathrm{al}}A_0G\beta\right)^{\phi_{\mathrm{mod}}^{2}}}
			\Delta_2
			\sum_{n=0}^{\infty}
			\frac{(-1)^n\Gamma(\eta)}
			{n!\,\Gamma(\eta-n)(1+n)^{\frac{\alpha-\phi_{\mathrm{mod}}^{2}}{\alpha}}}
			\\
			&\quad \times
			G_{1,2}^{2,0}\!\left(
			\frac{(1+n)h^{\alpha}}
			{\left(\beta A_0G h_{\mathrm{al}}\right)^{\alpha}}
			\,\middle|\,
			\begin{matrix}
				1 \\
				0,\; \dfrac{\alpha-\phi_{\mathrm{mod}}^{2}}{\alpha}
			\end{matrix}
			\right)
			+\delta(h)\Delta_3 ,
		\end{split}
	\end{equation}
	with $\Delta_2 = 1-\exp\,[-{\theta_{\mathrm{FoV}}^{2}}/({2\sigma_0^{2}})]$ and $\Delta_3 = \exp\,[-{\theta_{\mathrm{FoV}}^{2}}/({2\sigma_0^{2}})]$.
	Using the relationship for \ac{im/dd} \ac{ook} modulation scheme as $\gamma=\overline{\gamma}h^2$, the \ac{pdf} of the \ac{snr} can
	be expressed as \(f_{\gamma}(\gamma)=
	f_h\!\left(\sqrt{\gamma/\bar{\gamma}}\right)\left|dh/d\gamma\right|\) which results in
	\begin{equation}
		\label{eq:fgamma_final}
		\begin{split}
			f_{\gamma}(\gamma)
			&=
			\frac{\eta \phi_{\mathrm{mod}}^{2}\Delta_2}
			{2\Xi^{\phi_{\mathrm{mod}}^{2}}}
			\sum_{n=0}^{\infty}
			\frac{(-1)^n\Gamma(\eta)}
			{n!\,\Gamma(\eta-n)\,(1+n)^{\frac{\alpha-\phi_{\mathrm{mod}}^{2}}{\alpha}}}
			\\
			& \times
			\,\gamma^{\frac{\phi_{\mathrm{mod}}^{2}}{2}-1} G_{1,2}^{2,0}\!\left(
			\frac{(1+n)\gamma^{\alpha/2}}
			{\Xi^{\alpha}}
			\,\middle|\,
			\begin{matrix}
				1 \\
				0,\; \dfrac{\alpha-\phi_{\mathrm{mod}}^{2}}{\alpha}
			\end{matrix}
			\right)
			\\
			& +
			\frac{\Delta_3}{2\sqrt{\gamma\bar{\gamma}}}\,
			\delta\!\left(\sqrt{\frac{\gamma}{\bar{\gamma}}}\right),
		\end{split}
	\end{equation}
	where $\Xi = h_{\mathrm{al}}A_0G\beta\sqrt{\bar{\gamma}}$.
	\subsection{Outage Probability}
	The \ac{cdf} of the \ac{snr} \(\gamma\) is obtained as
	\begin{equation}
		\label{eq:Fgamma_def}
		F_{\gamma}(\gamma)=\int_{0}^{\gamma} f_{\gamma}(x)\,dx .
	\end{equation}
	
	Substituting \eqref{eq:fgamma_final} into \eqref{eq:Fgamma_def}, we obtain
	\begin{equation}
		\label{eq:Fgamma_final}
		\begin{split}
			&F_{\gamma}(\gamma)
			=
			\frac{\eta \phi_{\mathrm{mod}}^{2}\Delta_2}
			{2\Xi^{\phi_{\mathrm{mod}}^{2}}}
			\sum_{n=0}^{\infty}
			\frac{(-1)^n\Gamma(\eta)}
			{n!\,\Gamma(\eta-n)\,(1+n)^{\frac{\alpha-\phi_{\mathrm{mod}}^{2}}{\alpha}}}
			\\
			&\times
			\int_{0}^{\gamma}
			x^{\frac{\phi_{\mathrm{mod}}^{2}}{2}-1}
			G_{1,2}^{2,0}\!\left(
			\frac{(1+n)x^{\alpha/2}}
			{\Xi^{\alpha}}
			\,\middle|\,
			\begin{matrix}
				1 \\
				0,\; \dfrac{\alpha-\phi_{\mathrm{mod}}^{2}}{\alpha}
			\end{matrix}
			\right)\,dx
			\\
			& +
			\Delta_3
			\int_{0}^{\gamma}
			\frac{1}{2\sqrt{x\bar{\gamma}}}\,
			\delta\!\left(\sqrt{\frac{x}{\bar{\gamma}}}\right)\,dx .
		\end{split}
	\end{equation}
	For the first integral, we perform the change of variable \(t=x^{\alpha/2}\), then apply Eq.~(26) in \cite{Adamchik1990Algorithm}, and subsequently use Eq.~(07.34.16.0001.01) in \cite{Wolfram} to simplify the Meijer-\(G\) function. For the second integral, we use the change of variable \(z=\frac{1}{\sqrt{\bar{\gamma}}}x^{1/2}\). Accordingly, the \ac{cdf} of the \ac{snr} is obtained as
	\begin{equation}
		\label{eq:Fgamma_closed_form}
		\begin{split}
			&F_{\gamma}(\gamma)
			=
			\frac{\eta \phi_{\mathrm{mod}}^{2}\Delta_2}{\alpha}
			\sum_{n=0}^{\infty}
			\frac{(-1)^{n}\Gamma(\eta)}
			{n!\,\Gamma(\eta-n)\,(1+n)}
			\\
			&\times
			G_{2,3}^{2,1}\!\left(
			\frac{(1+n)\gamma^{\alpha/2}}
			{\Xi^{\alpha}}
			\,\middle|\,
			\begin{matrix}
				1,\; 1+\dfrac{\phi_{\mathrm{mod}}^{2}}{\alpha}
				\\[2pt]
				\dfrac{\phi_{\mathrm{mod}}^{2}}{\alpha},\; 1,\; 0
			\end{matrix}
			\right)+
			\Delta_3\,H\!\left(\sqrt{\frac{\gamma}{\bar{\gamma}}}\right),
		\end{split}
	\end{equation}
	where $H(.)$ is Heaviside function. Then, the \ac{op} is defined as
	\begin{equation}
		\label{eq:Pout_def}
		P_{\mathrm{out}}
		=
		\Pr(\gamma \leq \gamma_{\mathrm{th}})
		=
		F_{\gamma}(\gamma_{\mathrm{th}}),
	\end{equation}
	where \(\gamma_{\mathrm{th}}\) denotes the \ac{snr} threshold.
	
	By substituting \(\gamma=\gamma_{\mathrm{th}}\) into
	\eqref{eq:Fgamma_closed_form}, the \ac{op} is obtained as
	\begin{equation}
		\label{eq:Pout_final}
		\begin{split}
			&P_{\mathrm{out}}
			=
			\frac{\eta \phi_{\mathrm{mod}}^{2}\Delta_2}{\alpha}
			\sum_{n=0}^{\infty}
			\frac{(-1)^{n}\Gamma(\eta)}
			{n!\,\Gamma(\eta-n)\,(1+n)}
			\\
			&\times
			G_{2,3}^{2,1}\!\left(
			\frac{(1+n)\gamma_{\mathrm{th}}^{\alpha/2}}
			{\Xi^{\alpha}}
			\,\middle|\,
			\begin{matrix}
				1,\; 1+\dfrac{\phi_{\mathrm{mod}}^{2}}{\alpha}
				\\[2pt]
				\dfrac{\phi_{\mathrm{mod}}^{2}}{\alpha},\; 1,\; 0
			\end{matrix}
			\right)+
			\Delta_3\,H\!\left(\sqrt{\frac{\gamma_{\mathrm{th}}}{\bar{\gamma}}}\right).
		\end{split}
	\end{equation}
	
	\subsection{Average \ac{ber}}
	For an \ac{im/dd} \ac{ook} system, the \ac{ber} can be expressed as
	\begin{equation}
		\label{eq:Pb_def}
		P_b(e)=p(1)p(e|1)+p(0)p(e|0),
	\end{equation}
	where \(p(1)\) and \(p(0)\) denote the probabilities of transmitting
	bits ``1'' and ``0'', respectively, while \(p(e|1)\) and \(p(e|0)\)
	represent the corresponding conditional error probabilities. Assuming
	equiprobable bits, i.e., \(p(1)=p(0)=1/2\), and equal conditional error
		$P_b(e|\gamma)=Q\!\left(\sqrt{{\gamma}/{2}}\right)$,
	where \(Q(\sqrt{2}x)=\frac{1}{2}\operatorname{erfc}(x)\), \(Q(\cdot)\)
	is the Gaussian \(Q\)-function, and \(\operatorname{erfc}(\cdot)\) is
	the complementary error function.
	
	Accordingly, the average \ac{ber} is obtained by averaging
	\(P_b(e|\gamma)\) over the \ac{snr} \ac{pdf} \(f_{\gamma}(\gamma)\) as
	\begin{equation}
		\label{eq:Pb_avg_def}
		\overline{P}_b(e)=\int_{0}^{\infty} P_b(e|\gamma)\,f_{\gamma}(\gamma)\,d\gamma.
	\end{equation}
	
Expressing $\operatorname{erfc}(x)$ depending on Meijer-G function using Eq. (8.4.14.2) of \cite{Prudnikov1986IntegralsSeriesVol3}, the average \ac{ber} can be written as
	\begin{equation}
		\label{eq:Pb_avg_meijerg_compact}
		\begin{split}
			\overline{P}_b(e)
			&=
			\frac{\eta \phi_{\mathrm{mod}}^{2}\Delta_2}
			{4\sqrt{\pi}\,\Xi^{\phi_{\mathrm{mod}}^{2}}}
			\sum_{n=0}^{\infty}
			\frac{(-1)^n\Gamma(\eta)\,\mathcal{J}_n}
			{n!\,\Gamma(\eta-n)(1+n)^{\frac{\alpha-\phi_{\mathrm{mod}}^{2}}{\alpha}}}
			\\
			&\quad +
			\Delta_3
			\int_{0}^{\infty}
			\frac{1}{4\sqrt{\gamma\bar{\gamma}}}\,
			\delta\!\left(\sqrt{\frac{\gamma}{\bar{\gamma}}}\right)
			\operatorname{erfc}\!\left(\frac{\sqrt{\gamma}}{2}\right)d\gamma,
		\end{split}
	\end{equation}
	\begin{equation}
		\label{eq:J_n_def}
		\begin{split}
			\mathcal{J}_n
			&=
			\int_{0}^{\infty}
			\gamma^{\frac{\phi_{\mathrm{mod}}^{2}}{2}-1}
			G_{1,2}^{2,0}\!\left(
			\frac{\gamma}{4}
			\,\middle|\,
			\begin{matrix}
				1\\
				0,\; \dfrac{1}{2}
			\end{matrix}
			\right)
			\\
			&\quad \times
			G_{1,2}^{2,0}\!\left(
			\frac{(1+n)\gamma^{\alpha/2}}{\Xi^{\alpha}}
			\,\middle|\,
			\begin{matrix}
				1\\
				0,\; \dfrac{\alpha-\phi_{\mathrm{mod}}^{2}}{\alpha}
			\end{matrix}
			\right)
			\, d\gamma.
		\end{split}
	\end{equation}
	To solve integration in Eq. \eqref{eq:J_n_def}, we can use Eq. (07.34.21.0012.01) of \cite{Wolfram}. For the last integral in \eqref{eq:Pb_avg_meijerg_compact}, let
	\(z=\frac{1}{\sqrt{\bar{\gamma}}}\gamma^{1/2}\), and after some
	mathematical manipulations, the average \ac{ber} can be expressed as
	\begin{equation}
		\label{eq:Pb_avg_foxH}
		\begin{split}
			&\overline{P}_b(e)
			=\frac{\Delta_3}{2}+
			\frac{\eta \phi_{\mathrm{mod}}^{2}\Delta_2}
			{4\sqrt{\pi}\,
				\left(2\Delta_4\right)^{\phi_{\mathrm{mod}}^{2}}}
			\sum_{n=0}^{\infty}
			\frac{(-1)^n\Gamma(\eta)}
			{n!\,\Gamma(\eta-n)(1+n)^{\frac{\alpha-\phi_{\mathrm{mod}}^{2}}{\alpha}}}
			\\
			& \times
			H_{3,3}^{2,2}\!\left(
			\frac{1+n}
			{\Delta_4^{\alpha}}
			\,\middle|\,
			\begin{matrix}
				\left(\dfrac{2-\phi_{\mathrm{mod}}^{2}}{2},\dfrac{\alpha}{2}\right),
				\left(\dfrac{2-\phi_{\mathrm{mod}}^{2}}{2},\dfrac{\alpha}{2}\right),
				(1,1)
				\\[4pt]
				(0,1),
				\left(\dfrac{\alpha-\phi_{\mathrm{mod}}^{2}}{\alpha},1\right),
				\left(\dfrac{\phi_{\mathrm{mod}}^{2}}{2},\dfrac{\alpha}{2}\right)
			\end{matrix}
			\right) ,
		\end{split}
	\end{equation}
	where \(H_{p,q}^{m,n}(\cdot)\) denotes the Fox--\(H\) function and $\Delta_4=\beta A_0 G h_{\mathrm{al}}\sqrt{\bar{\gamma}}/2$. Using Eq. (2.1.5) of \cite{kilbas2004h}, \ac{ber} simplifies to
	\begin{equation}
		\label{eq:Pb_avg_foxH_simplified}
		\begin{split}
			&P_b(e)
			=\frac{1}{2}\Delta_3+
			\frac{\eta \phi_{\mathrm{mod}}^2 \Delta_2}{4\sqrt{\pi}}
			\sum_{n=0}^{\infty}
			\frac{(-1)^n\Gamma(\eta)}
			{n!\Gamma(\eta-n)(1+n)}
			\\
			&\times
			H_{3,3}^{2,2}\!\left(
			\frac{1+n}{\Delta_4^\alpha}
			\,\middle|\,
			\begin{matrix}
				\left(1,\dfrac{\alpha}{2}\right),
				\left(\dfrac{1}{2},\dfrac{\alpha}{2}\right),
				\left(\dfrac{\alpha+\phi_{\mathrm{mod}}^2}{\alpha},1\right)
				\\[6pt]
				\left(\dfrac{\phi_{\mathrm{mod}}^2}{\alpha},1\right),
				(1,1),
				\left(0,\dfrac{\alpha}{2}\right)
			\end{matrix}
			\right) .
		\end{split}
	\end{equation}
	\subsection{Ergodic Achievable-Rate Lower Bound}
	For the real-valued \ac{im/dd} optical-intensity channel, the input is nonnegative and its capacity depends on the imposed optical-intensity constraints. 
    Theorem~7 of \cite{lapidoth2009capacity} gives the following instantaneous achievable-rate lower bound. Averaging it over the channel \ac{snr} yields the \ac{ealb}
	\begin{equation}
		\label{eq:ergodic_capacity_def}
		\bar{C}_{\mathrm{LB}}
		=
		\frac{1}{2\ln(2)}
		\int_{0}^{\infty}
		\ln\!\left(1+\frac{e}{2\pi}\gamma\right)
		f_{\gamma}(\gamma)\,d\gamma.
	\end{equation}
	Substituting the PDF given in \eqref{eq:fgamma_final} into
	\eqref{eq:ergodic_capacity_def}, we obtain
	\begin{equation}
		\label{eq:ergodic_capacity_expanded}
		\begin{split}
			\bar{C}_{\mathrm{LB}}
			&=
			\frac{\eta \phi_{\mathrm{mod}}^{2}\Delta_2}
			{4\ln(2)\Xi^{\phi_{\mathrm{mod}}^{2}}}
			\sum_{n=0}^{\infty}
			\frac{(-1)^n\Gamma(\eta)\mathcal{I}_n}
			{n!\,\Gamma(\eta-n)(1+n)^{\frac{\alpha-\phi_{\mathrm{mod}}^{2}}{\alpha}}}
			\\
			&+
			\frac{\Delta_3}{2\ln(2)}
			\int_{0}^{\infty}
			\ln\!\left(1+\frac{e}{2\pi}\gamma\right)
			\frac{1}{2\sqrt{\gamma\bar{\gamma}}}
			\delta\!\left(\sqrt{\frac{\gamma}{\bar{\gamma}}}\right)
			\, d\gamma ,
		\end{split}
	\end{equation}
	where
	\begin{equation}
		\label{eq:ergodic_capacity_integral}
		\begin{split}
			\mathcal{I}_n=& \int_{0}^{\infty}
			\gamma^{\frac{\phi_{\mathrm{mod}}^{2}}{2}-1} \ln\!\left(1+\frac{e}{2\pi}\gamma\right)\\
			&\times 
			G_{1,2}^{2,0}\!\left(
			\frac{(1+n)\gamma^{\alpha/2}}
			{\Xi^{\alpha}}
			\,\middle|\,
			\begin{matrix}
				1 \\
				0,\; \dfrac{\alpha-\phi_{\mathrm{mod}}^{2}}{\alpha}
			\end{matrix}
			\right)
			\, d\gamma       
		\end{split}
	\end{equation}
	The last integral in Eq. \eqref{eq:ergodic_capacity_expanded} will result in zero. To solve integral in Eq. \eqref{eq:ergodic_capacity_integral}, we will use the identity \(\ln(1+x)=G_{2,2}^{1,2}\!\left(x \,\middle|\,
	\begin{matrix}
		1,1\\
		1,0
	\end{matrix}\right)\) \cite{Adamchik1990Algorithm}, then apply Eq. (07.34.21.0012.01) of \cite{Wolfram} and then apply Eq. (2.1.5) of \cite{kilbas2004h} to simplify Fox-H function, yielding the \ac{ealb} as
	\begin{equation}
		\label{eq:ergodic_capacity_foxH}
		\begin{split}
			&\bar{C}_{\mathrm{LB}}
			=
			\frac{\eta \phi_{\mathrm{mod}}^{2}\Delta_2}
			{4\ln(2)}
			\sum_{n=0}^{\infty}
			\frac{(-1)^n\Gamma(\eta)}
			{n!\,\Gamma(\eta-n)(1+n)}
			\\
			&\times
			H_{3,4}^{4,1}\!\left(
			\frac{1+n}
			{\left(\Xi\sqrt{\Delta_5}\right)^{\alpha}}
			\,\middle|\,
			\begin{matrix}
				\left(0,\dfrac{\alpha}{2}\right),
				\left(\dfrac{1}{2},\dfrac{\alpha}{2}\right),
				\left(1+\dfrac{\phi_{\mathrm{mod}}^{2}}{\alpha},1\right)
				\\[6pt]
				\left(\dfrac{\phi_{\mathrm{mod}}^{2}}{\alpha},1\right),
				(1,1),
				\left(0,\dfrac{\alpha}{2}\right),
				\left(0,\dfrac{\alpha}{2}\right)
			\end{matrix}
			\right),
		\end{split}
	\end{equation}
	where $\Delta_5=e/(2\pi)$.
	
	\section{Asymptotic Analysis for High \ac{snr} Regime} \label{sec:highsnr}
	
	In this subsection, high-\ac{snr} approximations are derived for the \ac{op}, average \ac{ber}, and \ac{ealb}. The derivation uses the small-argument behavior of the \ac{ew} turbulence CDF together with the mixed distribution induced by the \ac{aoa}/\ac{fov} loss.
	
	The instantaneous \ac{snr} is defined as
	\begin{equation}
		\gamma=\bar{\gamma}h^2,
	\end{equation}
	where the overall channel coefficient is given in Eq. \eqref{eq:overall_channel} as $h=h_{\mathrm{al}}h_{\mathrm{at}}h_{\mathrm{pl}}h_{\mathrm{af}}$.
	
	Following the \ac{aoa}/\ac{fov} model in Sec. III-C, the detection angle \(\theta_d\) is Rayleigh distributed \cite{Safi2020GroundHAP}, and the \ac{aoa}/\ac{fov} loss is modeled as
	\begin{equation}
		h_{\mathrm{af}}
		=
		\begin{cases}
			1, & \theta_d\le \theta_{\mathrm{FoV}},\\
			0, & \theta_d> \theta_{\mathrm{FoV}}.
		\end{cases}
	\end{equation}
	Therefore,
	\begin{equation}
		\Delta_2
		=
		\Pr(\theta_d\le \theta_{\mathrm{FoV}})
		=
		1-\exp\left(
		-\frac{\theta_{\mathrm{FoV}}^2}{2\sigma_0^2}
		\right),
	\end{equation}
	\begin{equation}
		\Delta_3
		=
		\Pr(\theta_d> \theta_{\mathrm{FoV}})
		=
		\exp\left(
		-\frac{\theta_{\mathrm{FoV}}^2}{2\sigma_0^2}
		\right).
	\end{equation}
	Thus, \(h_{\mathrm{af}}\) introduces a point mass at zero channel gain with probability \(\Delta_3\), while the nonblocked channel occurs with probability \(\Delta_2\). Let the nonblocked composite channel be
	\begin{equation}
		h_{\mathrm{ag}}=h_{\mathrm{al}}h_{\mathrm{at}}h_{\mathrm{pl}} .
	\end{equation}
	Then, for \(x>0\), the CDF of the overall channel is written as
	\begin{equation}
		F_h(x)
		=
		\Delta_3
		+
		\Delta_2 F_{h_{\mathrm{ag}}}(x).
	\end{equation}
	
	The \ac{ew} CDF of the turbulence-induced fading is \cite{barrios2012exponentiated}
	\begin{equation}
		F_{h_{\mathrm{at}}}(x)
		=
		\left[
		1-\exp\left(
		-\left(\frac{x}{\beta}\right)^\alpha
		\right)
		\right]^\eta .
	\end{equation}
	In the high-\ac{snr} regime, outage and error events are dominated by the small-channel-gain region. Hence, as \(x\to 0\),
	\begin{equation}
		1-\exp\left(
		-\left(\frac{x}{\beta}\right)^\alpha
		\right)
		\simeq
		\left(\frac{x}{\beta}\right)^\alpha,
	\end{equation}
	and therefore
	\begin{equation}
		F_{h_{\mathrm{at}}}(x)
		\simeq
		\left({x}/{\beta}\right)^{\alpha\eta}.
	\end{equation}
	Using the Beckmann-based \ac{pe} approximation \cite{BoludaRuiz2016Beckmann}, and using the same small-angle approximation \(\cos(\theta_d)\simeq 1\), the \ac{pe} fading coefficient has the approximate \ac{pdf}
	\begin{equation}
		f_{h_{\mathrm{pl}}}(u)
		=
		\frac{\phi_{\mathrm{mod}}^2}
		{(A_0G)^{\phi_{\mathrm{mod}}^2}}
		u^{\phi_{\mathrm{mod}}^2-1},
		\qquad
		0\le u\le A_0G .
	\end{equation}
	For notational compactness, let's define $\kappa=\alpha\eta$. Then, the high-\ac{snr} \ac{cdf} of the overall channel is
	\begin{equation}
		F_h(x)
		\simeq
		\Delta_3
		+
		\Delta_2
		\int_0^{A_0G}
		F_{h_{\mathrm{at}}}
		\left(
		\frac{x}{h_{\mathrm{al}}u}
		\right)
		f_{h_{\mathrm{pl}}}(u)
		\,du .
	\end{equation}
	Substituting the small-argument \ac{ew} approximation gives
	\begin{equation}
		\begin{split}
			F_h(x)
			&\simeq
			\Delta_3
			+
			\Delta_2
			\int_0^{A_0G}
			\left(
			\frac{x}{h_{\mathrm{al}}\beta u}
			\right)^\kappa
			\frac{\phi_{\mathrm{mod}}^2}
			{(A_0G)^{\phi_{\mathrm{mod}}^2}}
			u^{\phi_{\mathrm{mod}}^2-1}
			\,du .
		\end{split}
	\end{equation}
	For \(\kappa<\phi_{\mathrm{mod}}^2\), which is the regime considered in the numerical results, this integral yields
	\begin{equation}
		F_h(x)
		\simeq
		\Delta_3
		+
		\Delta_2
		\frac{\phi_{\mathrm{mod}}^2}
		{\phi_{\mathrm{mod}}^2-\kappa}
		\left(
		\frac{x}
		{h_{\mathrm{al}}A_0G\beta}
		\right)^\kappa .
	\end{equation}
	
	Since \(\gamma=\bar{\gamma}h^2\), the high-\ac{snr} CDF of \(\gamma\) is
	\begin{equation}
		F_\gamma(\gamma)
		\simeq
		\Delta_3
		+
		\Delta_2
		\frac{\phi_{\mathrm{mod}}^2}
		{\phi_{\mathrm{mod}}^2-\kappa}
		\left(
		\frac{\sqrt{\gamma}}
		{h_{\mathrm{al}}A_0G\beta\sqrt{\bar{\gamma}}}
		\right)^\kappa .
	\end{equation}
	
	\subsubsection{Outage Probability}
	
	The \ac{op} is defined as
	\begin{equation}
		P_{\mathrm{out}}
		=
		F_\gamma(\gamma_{\mathrm{th}}).
	\end{equation}
	Therefore, the high-\ac{snr} \ac{op} is
	\begin{equation}
		P_{\mathrm{out}}^{\infty}
		\simeq
		\Delta_3
		+
		\Delta_2
		\frac{\phi_{\mathrm{mod}}^2}
		{\phi_{\mathrm{mod}}^2-\kappa}
		\left(
		\frac{\sqrt{\gamma_{\mathrm{th}}}}
		{h_{\mathrm{al}}A_0G\beta}
		\right)^\kappa \bar{\gamma}^{-\kappa/2}.
	\end{equation}
	
	Because of the \ac{aoa}/\ac{fov} blocking event, the \ac{op} does not vanish at high \ac{snr}; instead, it approaches the floor \(\Delta_3\). The nonblocked component decays with diversity order
	\begin{equation}
		G_d=\kappa/2=\alpha\eta/2.
	\end{equation}

	More generally, the decay exponent of the nonblocked composite channel is determined by the smaller of the \ac{ew} small-gain exponent and the pointing-loss exponent. Therefore,
	\begin{equation}
		G_d^{\mathrm{nb}}=\frac{1}{2}\min\!\left(\alpha\eta,\phi_{\mathrm{mod}}^2\right).
	\end{equation}
	For the regime $\kappa=\alpha\eta<\phi_{\mathrm{mod}}^2$ considered here, define
	\begin{equation}
		A_{\mathrm{out}}=\Delta_2\frac{\phi_{\mathrm{mod}}^2}{\phi_{\mathrm{mod}}^2-\kappa}\left(\frac{\sqrt{\gamma_{\mathrm{th}}}}{h_{\mathrm{al}}A_0G\beta}\right)^\kappa.
	\end{equation}
	Writing the floor-subtracted asymptote as
	\begin{equation}
		P_{\mathrm{out}}^\infty-\Delta_3\simeq\left(G_{c,\mathrm{out}}\bar{\gamma}\right)^{-G_d^{\mathrm{nb}}},
	\end{equation}
	gives the effective outage coding gain
	\begin{equation}
		G_{c,\mathrm{out}}=A_{\mathrm{out}}^{-2/\kappa}.
	\end{equation}
	The total \ac{op} has zero conventional diversity order because it approaches the nonzero floor $\Delta_3$; hence, $G_{c,\mathrm{out}}$ characterizes only its decaying nonblocked component. In physical terms, \ac{aoc} changes $\alpha$ and hence the turbulence-controlled diversity order, whereas atmospheric attenuation, aperture collection, the \ac{ew} scale parameter, and the pointing parameters enter the effective coding gain. The diversity improvement obtained by increasing the correction order is ultimately limited by the pointing-loss exponent $\phi_{\mathrm{mod}}^2/2$.
	
	\subsubsection{Average \ac{ber}}
	For the considered \ac{im/dd} \ac{ook} system, the conditional \ac{ber} is 
	\begin{equation}
		P_b(e|\gamma)
		=
		\frac{1}{2}
		\operatorname{erfc}
		\left(
		\frac{\sqrt{\gamma}}{2}
		\right).
	\end{equation}
	The continuous part of the high-\ac{snr} \ac{cdf} can be written as
	\begin{equation}
		F_\gamma(\gamma)-\Delta_3
		\simeq
		K_\gamma \gamma^{\kappa/2}\bar{\gamma}^{-\kappa/2},
	\end{equation}
	where
	\begin{equation}
		K_\gamma
		=
		\Delta_2
		\frac{\phi_{\mathrm{mod}}^2}
		{\phi_{\mathrm{mod}}^2-\kappa}
		\left(
		\frac{1}
		{h_{\mathrm{al}}A_0G\beta}
		\right)^\kappa .
	\end{equation}
	Hence, the continuous part of the high-\ac{snr} \ac{pdf} is
	\begin{equation}
		f_\gamma^\infty(\gamma)
		\simeq
		K_\gamma
		\frac{\kappa}{2}
		\gamma^{\kappa/2-1}
		\bar{\gamma}^{-\kappa/2}.
	\end{equation}
	
	The average \ac{ber} is
	\begin{equation}
		P_b(e)
		=
		\int_0^\infty
		P_b(e|\gamma)f_\gamma(\gamma)\,d\gamma .
	\end{equation}
	Since \(h_{\mathrm{af}}=0\) gives \(\gamma=0\), the blocked component contributes \(\Delta_3/2\). Therefore,
	\begin{equation}
		\begin{split}
			P_b^\infty(e)
			&\simeq
			\frac{\Delta_3}{2}
			+
			K_\gamma
			\frac{\kappa}{4}
			\bar{\gamma}^{-\kappa/2}
			\int_0^\infty
			\gamma^{\kappa/2-1}
			\operatorname{erfc}
			\left(
			\frac{\sqrt{\gamma}}{2}
			\right)
			d\gamma .
		\end{split}
	\end{equation}
	Using Eq. (2.8.2-1) of \cite{prudnikov1992integrals}, the integral identity becomes
	\begin{equation}
		\int_0^\infty
		\gamma^{\kappa/2-1}
		\operatorname{erfc}
		\left(
		\frac{\sqrt{\gamma}}{2}
		\right)
		d\gamma
		=
		\frac{
			2^{\kappa+1}
			\Gamma\left(\frac{\kappa+1}{2}\right)
		}
		{\kappa\sqrt{\pi}},
	\end{equation}
	the high-\ac{snr} average \ac{ber} becomes
	\begin{equation}
		\begin{split}
			P_b^\infty(e)
			\simeq &
			\frac{\Delta_3}{2}
			+
			\Delta_2
			\frac{\phi_{\mathrm{mod}}^2}
			{\phi_{\mathrm{mod}}^2-\kappa}
			\left(
			\frac{1}
			{h_{\mathrm{al}}A_0G\beta}
			\right)^\kappa \\
			&\times 
			\frac{
				2^{\kappa-1}
				\Gamma\left(\frac{\kappa+1}{2}\right)
			}
			{\sqrt{\pi}}
			\bar{\gamma}^{-\kappa/2}
		\end{split}
	\end{equation}
	where \(\kappa=\alpha\eta\). Hence, the average \ac{ber} also has a floor equal to \(\Delta_3/2\). Defining
	\begin{equation}
		\begin{split}
			A_{\mathrm{BER}}&=\Delta_2\frac{\phi_{\mathrm{mod}}^2}{\phi_{\mathrm{mod}}^2-\kappa}\left(\frac{1}{h_{\mathrm{al}}A_0G\beta}\right)^\kappa\\
			&\quad\times\frac{2^{\kappa-1}\Gamma\!\left((\kappa+1)/2\right)}{\sqrt{\pi}}.
		\end{split}
	\end{equation}
	the floor-subtracted average \ac{ber} can be written as $P_b^\infty(e)-\Delta_3/2\simeq\left(G_{c,\mathrm{BER}}\bar{\gamma}\right)^{-G_d^{\mathrm{nb}}}$, where
	\begin{equation}
		G_{c,\mathrm{BER}}=A_{\mathrm{BER}}^{-2/\kappa}.
	\end{equation}
	Thus, the floor-subtracted outage and average \ac{ber} have the same diversity order, while their effective coding gains differ. The total average \ac{ber}, like the total outage probability, has zero conventional diversity order because of its nonzero blocking floor.
	
	\subsubsection{Ergodic Achievable-Rate Lower Bound}
	For \ac{im/dd} detection under the stated average optical-intensity constraint, the \ac{ealb} is \cite{lapidoth2009capacity}
	\begin{equation}
		\bar{C}_{\mathrm{LB}}
		=
		\frac{1}{2\ln 2}
		\mathbb{E}
		\left[
		\ln\left(
		1+\frac{e}{2\pi}\gamma
		\right)
		\right].
	\end{equation}
	For \(\bar{\gamma}\gg 1\),
	\begin{equation}
		\ln\left(
		1+\frac{e}{2\pi}\gamma
		\right)
		\simeq
		\ln\left(
		\frac{e}{2\pi}\gamma
		\right).
	\end{equation}
	The blocked component, for which \(h_{\mathrm{af}}=0\), contributes zero achievable rate. Hence,
	\begin{equation}
		\begin{split}
			\bar{C}_{\mathrm{LB}}^{\infty}
			&\simeq
			\frac{\Delta_2}{2\ln 2}
			\mathbb{E}
			\left[
			\ln\left(
			\frac{e}{2\pi}
			\bar{\gamma}
			h_{\mathrm{al}}^2
			h_{\mathrm{at}}^2
			h_{\mathrm{pl}}^2
			\right)
			\right].
		\end{split}
	\end{equation}
	This can be expanded as
	\begin{equation}
		\begin{split}
			\bar{C}_{\mathrm{LB}}^{\infty}
			&\simeq
			\frac{\Delta_2}{2\ln 2}
			\bigg[
			\ln\left(
			\frac{e}{2\pi}
			\bar{\gamma}
			h_{\mathrm{al}}^2
			\right)
			+
			2\mathbb{E}\{\ln h_{\mathrm{at}}\}
			+
			2\mathbb{E}\{\ln h_{\mathrm{pl}}\}
			\bigg].
		\end{split}
	\end{equation}
	
	For the \ac{pe} fading coefficient,
	\begin{equation}
		\mathbb{E}\{\ln h_{\mathrm{pl}}\}
		=
		\ln(A_0G)-{1}/{\phi_{\mathrm{mod}}^2}.
	\end{equation}
	For the \ac{ew} turbulence coefficient, using the Newton generalized binomial expansion,
	\begin{equation}
		\mathbb{E}\{\ln h_{\mathrm{at}}\}
		=
		\ln\beta
		-
		\frac{\eta}{\alpha}
		\sum_{n=0}^{\infty}
		\frac{
			(-1)^n\Gamma(\eta)
		}
		{
			n!\Gamma(\eta-n)(1+n)
		}
		\left[
		\gamma_E+\ln(1+n)
		\right],
	\end{equation}
	where $\gamma_E\approx0.5772$ is the Euler--Mascheroni constant \cite{favano2025non}.
	
	Then, the corrected high-\ac{snr} \ac{ealb} is
	\begin{equation}
		\begin{split}
			\bar{C}_{\mathrm{LB}}^{\infty}
			&\simeq
			\frac{\Delta_2}{2\ln 2}
			\bigg[
			\ln\left(
			\frac{e}{2\pi}
			\bar{\gamma}
			(h_{\mathrm{al}}A_0G\beta)^2
			\right)
			-
			\frac{2}{\phi_{\mathrm{mod}}^2}
			\\
			&-
			\frac{2\eta}{\alpha}
			\sum_{n=0}^{\infty}
			\frac{
				(-1)^n\Gamma(\eta)
			}
			{
				n!\Gamma(\eta-n)(1+n)
			}
			\left[
			\gamma_E+\ln(1+n)
			\right]
			\bigg].
		\end{split}
	\end{equation}
	
	\subsection{Computational Complexity}
	The special-function and infinite-series forms above provide analytical characterizations of the composite channel, but their direct symbolic evaluation is not required for the numerical results. The \ac{op} is evaluated using one-dimensional Gauss--Legendre quadrature over the pointing-loss variable, with complexity $O(Q_{\mathrm{pl}})$ per operating point. The average \ac{ber} and \ac{ealb} use two-dimensional quadrature over the \ac{ew} turbulence and pointing-loss variables and therefore require $O(Q_{\mathrm{EW}}Q_{\mathrm{pl}})$ operations per operating point. The quadrature nodes, weights, and transformed channel samples can be precomputed and reused in parameter sweeps. The short series used to obtain the \ac{ew} scale parameter and the adaptively terminated logarithmic-moment series have complexity linear in their retained term counts. After these quantities are calculated, the high-\ac{snr} outage and \ac{ber} expressions have $O(1)$ evaluation complexity per \ac{snr} value. Monte Carlo simulation scales linearly with the number of channel realizations and is used only for offline verification. Consequently, the proposed analysis does not impose real-time processing at the transceiver; it is intended as an offline link-analysis and design tool. The hardware cost and real-time control complexity of a specific \ac{ao} implementation are not represented by the reduced-order statistical model.

	\section{Results and Discussion} \label{sec:results_discussions}
	
	Unless otherwise stated, the numerical results are obtained using the simulation parameters listed in Table~\ref{tab:simulation_parameters}. Both \ac{ul} and \ac{dl} transmissions are considered.

	The \ac{mc} simulations generate the channel coefficients according to the same adopted composite statistical model used in the analytical derivations, including the the turbulence, pointing-loss, and \ac{aoa}/\ac{fov}-blocking random variables and combine them with the deterministic atmospheric attenuation according to the adopted composite channel model. Therefore, agreement between the analytical and \ac{mc} results verifies the derived transformations, performance expressions, series evaluation, and numerical implementation under the adopted model. The accuracy of that approximation relative to the original Beckmann model was assessed independently in \cite{BoludaRuiz2016Beckmann}. The high-\ac{snr} expressions are shown separately and approach the analytical results only in their intended asymptotic region.\\

	\begin{table}[!t]
		\centering
		\caption{Simulation Parameters Used in the Numerical Results}
		\label{tab:simulation_parameters}
		\renewcommand{\arraystretch}{1.05}
		\setlength{\tabcolsep}{2.5pt}
		\scriptsize
		\begin{tabular}{l c | l c}
			\hline
			\textbf{Parameter} & \textbf{Value} & \textbf{Parameter} & \textbf{Value} \\
			\hline
			\hline 
			$H_H$ & $20~\mathrm{km}$ &
			$h_{\mathrm{GS}}$ & $20~\mathrm{m}$ \\
			
			$\zeta$ & $40^\circ$ &
			$\bar{\gamma}$ & $40~\mathrm{dB}$ \\
			
			$\gamma_{\mathrm{th}}$ & $5~\mathrm{dB}$ &
			$N$ & $\{3,5,7,10,20,35\}$ \\
			
			$\lambda$ & $1550~\mathrm{nm}$ &
			$D_G$ & $0.20~\mathrm{m}$ \\
			
			$D_t$ & $0.10~\mathrm{m}$ &
			$W_0$ & $0.05~\mathrm{m}$ \\
			
			$F_0$ & $\infty$ &
			$V$ & $23~\mathrm{km}$ \\
			
			$H_0$ & $1.2~\mathrm{km}$ &
			$\sigma_0$ & $5~\mathrm{mrad}$ \\
			
			$\theta_{\mathrm{FoV}}$ & $20~\mathrm{mrad}$ &
			$w$ & $21~\mathrm{m/s}$ \\
			
			$A= C_n^2(0)$ & $1.7\times10^{-14}~\mathrm{m}^{-2/3}$ &
			$\mu_x$ & $5~\mathrm{mm}$ \\
			
			$\mu_y$ & $5~\mathrm{mm}$ &
			$\sigma_x=\sigma_y$ & $0.01~\mathrm{m}$ \\
			\hline
		\end{tabular}
	\end{table}

	\begin{figure}[!t]
		\centering
		
		{
			\includegraphics[width=\columnwidth]{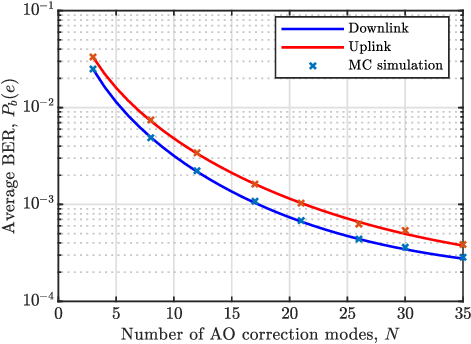}
			\label{fig:ber_versus_n}
		}
		
		\vspace{1mm}
		
		{
			\includegraphics[width=\columnwidth]{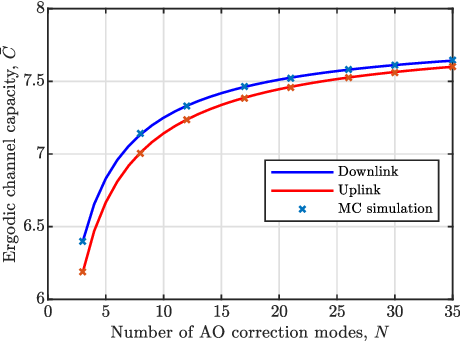}
			\label{fig:capacity_versus_n}
		}
		
		\caption{\ac{ber} and \ac{ealb} performance versus the number of corrected \ac{ao} modes.}
		\label{fig:ber_capacity_versus_n}
	\end{figure}
	Fig.~\ref{fig:ber_capacity_versus_n} jointly shows the average \ac{ber} and \ac{ealb} versus the number of corrected \ac{ao} modes, $N$. Increasing $N$ reduces the residual turbulence-induced irradiance fluctuation, thereby improving both reliability and the achievable rate. In the \ac{dl}, the \ac{ber} decreases from about $2.5\times10^{-2}$ at $N=3$ to $3.4\times10^{-3}$, $1.1\times10^{-3}$, and $2.5\times10^{-4}$ at $N=8$, $17$, and $35$, respectively; the corresponding \ac{ul} values are approximately $3.2\times10^{-2}$, $5.8\times10^{-3}$, $1.6\times10^{-3}$, and $3.5\times10^{-4}$. The \ac{ealb} follows the opposite trend and gradually saturates: from $N=3$ to $35$, it increases from about $3.2$ to $3.83$ bits/channel use in the \ac{dl} and from $3.1$ to $3.8$ bits/channel use in the \ac{ul}. Thus, the main \ac{ao} gain is obtained at low-to-moderate correction orders, while additional modes provide diminishing returns.
	
	\begin{figure}[!t]
		\centering
		\includegraphics[width=\columnwidth]{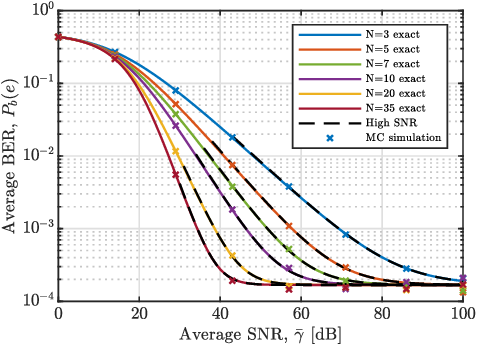}
		\caption{Average \ac{ber} versus average \ac{snr} for the \ac{dl} case.}
		\label{fig:ber_versus_average_snr_downlink}
	\end{figure}
	
	Fig.~\ref{fig:ber_versus_average_snr_downlink} plots the \ac{dl} average \ac{ber} versus $\bar{\gamma}$. Higher-order \ac{aoc} shifts the curves leftward, reducing the required \ac{snr} for a target error rate. At $\bar{\gamma}=43$ dB, the \ac{ber} decreases from about $1.8\times10^{-2}$ for $N=3$ to $7\times10^{-3}$, $3.5\times10^{-3}$, $1.6\times10^{-3}$, $4.5\times10^{-4}$, and $2\times10^{-4}$ for $N=5$, $7$, $10$, $20$, and $35$, respectively. Equivalently, for $P_b(e)\approx10^{-3}$, the required $\bar{\gamma}$ decreases from about $69$ dB for $N=3$ to $34$ dB for $N=35$, yielding an \ac{snr} saving of nearly $35$ dB. The analytical curves approach the high-\ac{snr} approximations only in the asymptotic region, as expected.
	
	\begin{figure}[!t]
		\centering
		\includegraphics[width=\columnwidth]{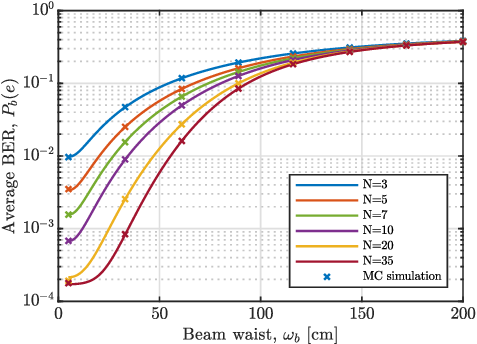}
		\caption{Average \ac{ber} versus beam waist for the \ac{ul} case.}
		\label{fig:ber_versus_beam_waist_uplink}
	\end{figure}
	Fig.~\ref{fig:ber_versus_beam_waist_uplink} shows the \ac{ul} average \ac{ber} versus beam waist. Smaller beam waists provide lower error rates, especially with higher-order correction: at $\omega_b=5$ cm, the \ac{ber} decreases from about $9\times10^{-3}$ for $N=3$ to $1.5\times10^{-4}$ for $N=35$, while at $\omega_b=30$ cm it decreases from $4.5\times10^{-2}$ to below $10^{-3}$. As $\omega_b$ increases, the received intensity concentration weakens and the \ac{ber} rises for all curves. Above roughly $120$ cm, the curves converge toward $2\times10^{-1}$--$4\times10^{-1}$, indicating that geometric spreading dominates over the residual turbulence reduction.
	
	\begin{figure}[!t]
		\centering
		\includegraphics[width=\columnwidth]{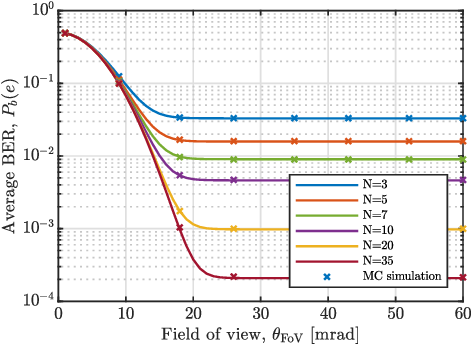}
		\caption{Average \ac{ber} versus receiver \ac{fov} for the \ac{ul} case.}
		\label{fig:ber_versus_field_of_view_uplink}
	\end{figure}
	Fig.~\ref{fig:ber_versus_field_of_view_uplink} presents the \ac{ul} average \ac{ber} versus receiver \ac{fov}. For a very narrow acceptance angle, $\theta_{\mathrm{FoV}}=2$ mrad, all curves give a high \ac{ber} of about $4.5\times10^{-1}$; increasing $\theta_{\mathrm{FoV}}$ to $9$ mrad lowers it to nearly $10^{-1}$. For $\theta_{\mathrm{FoV}}\geq25$ mrad, the curves reach mode-dependent floors of approximately $3\times10^{-2}$, $1.5\times10^{-2}$, $9\times10^{-3}$, $4.5\times10^{-3}$, $9\times10^{-4}$, and $2\times10^{-4}$ for $N=3$, $5$, $7$, $10$, $20$, and $35$, respectively. Hence, once the \ac{fov} is sufficiently wide, the performance is limited mainly by residual turbulence, and higher-order \ac{ao} provides more than two orders of magnitude \ac{ber} reduction.
	
	\begin{figure}[!t]
		\centering
		\includegraphics[width=\columnwidth]{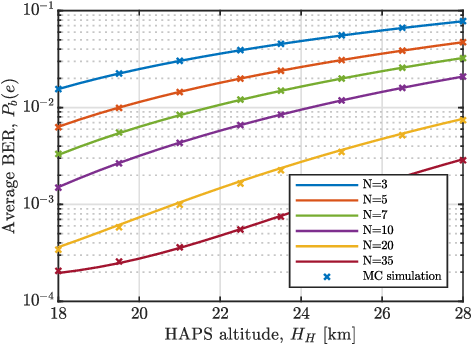}
		\caption{Average \ac{ber} versus \ac{haps} altitude for the \ac{dl} case.}
		\label{fig:ber_versus_haps_altitude_downlink}
	\end{figure}
	Fig.~\ref{fig:ber_versus_haps_altitude_downlink} shows that the \ac{dl} \ac{ber} increases monotonically with \ac{haps} altitude because of the longer propagation distance and stronger accumulated attenuation/fading. For $N=3$, the \ac{ber} rises from about $1.5\times10^{-2}$ at $H_H=18$ km to $7.5\times10^{-2}$ at $28$ km. With $N=35$, the corresponding values are much lower, namely about $2\times10^{-4}$ and $3\times10^{-3}$. Thus, even at the largest altitude, higher-order \ac{ao} yields more than one order of magnitude improvement relative to \ac{ttc}.
	
	\begin{figure}[!t]
		\centering
		\includegraphics[width=\columnwidth]{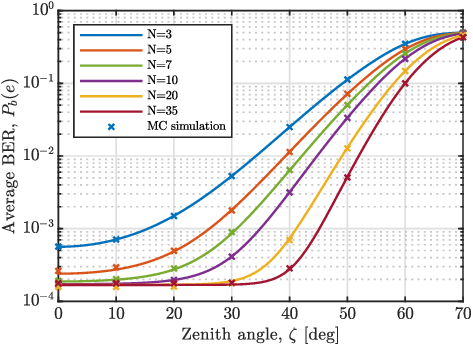}
		\caption{Average \ac{ber} versus zenith angle for the \ac{dl} case.}
		\label{fig:ber_versus_zenith_downlink}
	\end{figure}
	Fig.~\ref{fig:ber_versus_zenith_downlink} demonstrates the strong \ac{dl} sensitivity to zenith angle. For $N=3$, the \ac{ber} increases from about $5\times10^{-4}$ at $\zeta=0^\circ$ to $5\times10^{-3}$ at $30^\circ$, $2.5\times10^{-2}$ at $40^\circ$, and nearly $4.5\times10^{-1}$ at $70^\circ$, owing to the increased slant-path length. Higher-order \ac{ao} shifts this degradation to larger angles; at $\zeta=40^\circ$, for instance, the \ac{ber} decreases from $2.5\times10^{-2}$ for $N=3$ to about $2\times10^{-4}$ for $N=35$. Near $\zeta=70^\circ$, however, all curves enter the high-\ac{ber} region, showing that severe slant-path loss eventually limits the \ac{ao} gain.
	
	\begin{figure}[!t]
		\centering
		\includegraphics[width=\columnwidth]{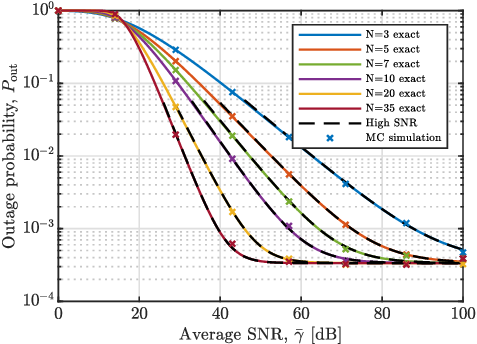}
		\caption{\ac{op} versus average \ac{snr} for the \ac{ul} case.}
		\label{fig:outage_versus_average_snr_uplink}
	\end{figure}
	Fig.~\ref{fig:outage_versus_average_snr_uplink} shows the \ac{ul} \ac{op} versus $\bar{\gamma}$. At $\bar{\gamma}=43$ dB, $P_{\mathrm{out}}$ decreases from about $7\times10^{-2}$ for $N=3$ to $3\times10^{-2}$, $2\times10^{-2}$, $9\times10^{-3}$, $1.5\times10^{-3}$, and $6\times10^{-4}$ for $N=5$, $7$, $10$, $20$, and $35$, respectively. At a fixed outage of $10^{-3}$, the required $\bar{\gamma}$ is approximately $86$, $71$, $58$, $56$, $43$, and $41$ dB for the same $N$ values, so $N=35$ provides about $45$ dB \ac{snr} saving compared with \ac{ttc}. The high-\ac{snr} curves accurately track the exact results in the asymptotic region.
	
	\begin{figure}[!t]
		\centering
		\includegraphics[width=\columnwidth]{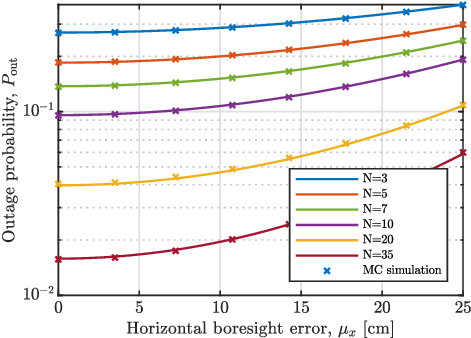}
		\caption{\ac{op} versus horizontal boresight error for the \ac{ul} case.}
		\label{fig:outage_versus_boresight_error_uplink}
	\end{figure}
	Fig.~\ref{fig:outage_versus_boresight_error_uplink} presents the \ac{ul} \ac{op} versus horizontal boresight error. Increasing $\mu_x$ raises $P_{\mathrm{out}}$ for all correction orders, showing that deterministic pointing bias remains important after turbulence mitigation. At $\mu_x=0$ cm, $P_{\mathrm{out}}$ is approximately $3.8\times10^{-1}$, $2.3\times10^{-1}$, $1.4\times10^{-1}$, $9\times10^{-2}$, $4.2\times10^{-2}$, and $1.5\times10^{-2}$ for $N=3$, $5$, $7$, $10$, $20$, and $35$, respectively. At $\mu_x=25$ cm, these values increase to about $5.4\times10^{-1}$, $4.1\times10^{-1}$, $3.0\times10^{-1}$, $2.4\times10^{-1}$, $1.1\times10^{-1}$, and $5\times10^{-2}$, so higher-order correction still gives roughly one order of magnitude improvement under severe misalignment.
	
	\begin{figure}[!t]
		\centering
		\includegraphics[width=\columnwidth]{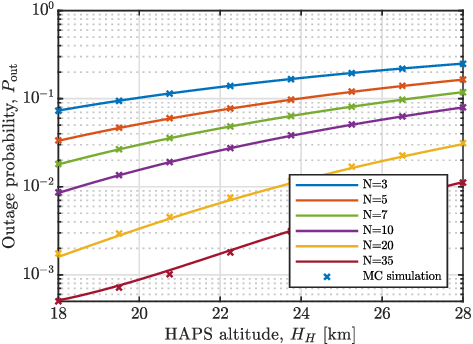}
		\caption{\ac{op} versus \ac{haps} altitude for the \ac{ul} case.}
		\label{fig:outage_versus_haps_altitude_uplink}
	\end{figure}
	Fig.~\ref{fig:outage_versus_haps_altitude_uplink} shows that \ac{ul} outage increases with \ac{haps} altitude because of the longer path length. For $N=3$, $P_{\mathrm{out}}$ grows from about $7.5\times10^{-2}$ at $H_H=18$ km to $2.5\times10^{-1}$ at $28$ km, whereas for $N=35$ it increases only from about $5\times10^{-4}$ to $10^{-2}$. Around the nominal altitude, $H_H=23.75$ km, the \acp{op} are approximately $1.6\times10^{-1}$, $9.5\times10^{-2}$, $6\times10^{-2}$, $3.8\times10^{-2}$, $1.1\times10^{-2}$, and $3\times10^{-3}$ for $N=3$, $5$, $7$, $10$, $20$, and $35$, respectively, corresponding to nearly two orders of magnitude improvement.
	
	\begin{figure}[!t]
		\centering
		\includegraphics[width=\columnwidth]{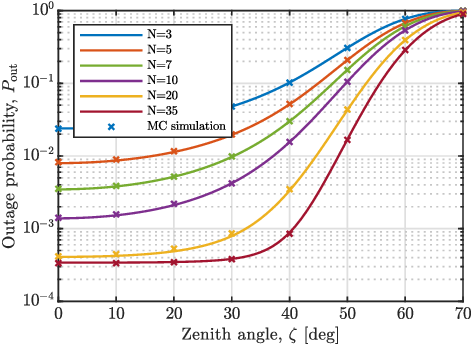}
		\caption{\ac{op} versus zenith angle for the \ac{ul} case.}
		\label{fig:outage_versus_zenith_uplink}
	\end{figure}
	Fig.~\ref{fig:outage_versus_zenith_uplink} illustrates the \ac{ul} outage sensitivity to zenith angle. For $N=3$, $P_{\mathrm{out}}$ increases from about $2.4\times10^{-2}$ at $\zeta=0^\circ$ to nearly $10^{-1}$ at $40^\circ$ and $7.5\times10^{-1}$ at $60^\circ$. Higher-order \ac{ao} shifts this degradation to larger angles: at $\zeta=40^\circ$, $P_{\mathrm{out}}$ decreases from about $10^{-1}$ for $N=3$ to $8\times10^{-4}$ for $N=35$, and at $\zeta=50^\circ$ from about $3\times10^{-1}$ to $1.5\times10^{-2}$. As $\zeta$ approaches $70^\circ$, all curves converge toward the outage-limited region, indicating that severe slant-path loss dominates.
	
	\begin{figure}[!t]
		\centering
		\includegraphics[width=\columnwidth]{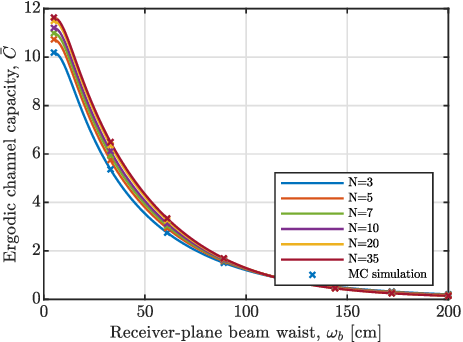}
		\caption{\ac{ealb} versus receiver-plane beam waist for the \ac{dl} case.}
		\label{fig:capacity_versus_beam_waist_downlink}
	\end{figure}
	Fig.~\ref{fig:capacity_versus_beam_waist_downlink} shows the \ac{dl} \ac{ealb} versus receiver-plane beam waist. The achievable-rate lower bound decreases as $\omega_b$ increases because a wider received beam lowers the collected optical intensity. For $N=3$, it drops from about $5.1$ bits/channel use at $\omega_b=5$ cm to $2.7$ bits/channel use at $33$ cm, $1.35$ bits/channel use at $61$ cm, and below $0.15$ bits/channel use beyond $170$ cm. The \ac{ao} gain is most visible for small-to-moderate beam waists: $N=35$ improves the lower bound from $5.1$ to $5.85$ bits/channel use at $5$ cm and from $2.7$ to $3.25$ bits/channel use at $33$ cm, while the curves converge for large $\omega_b$.
	
	\begin{figure}[!t]
		\centering
		\includegraphics[width=\columnwidth]{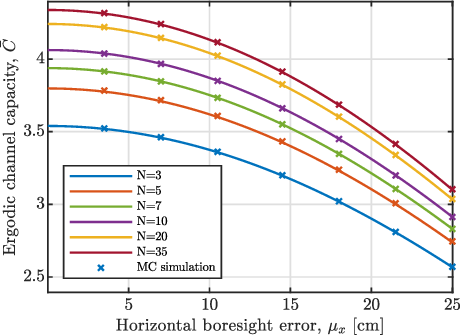}
		\caption{\ac{ealb} versus horizontal boresight error for the \ac{dl} case.}
		\label{fig:capacity_versus_boresight_error_downlink}
	\end{figure}
	Fig.~\ref{fig:capacity_versus_boresight_error_downlink} presents the \ac{dl} \ac{ealb} versus horizontal boresight error. The lower bound decreases monotonically with $\mu_x$ because misalignment reduces the collected optical power. For $N=3$, it decreases from about $1.78$ bits/channel use at $\mu_x=0$ cm to $1.6$ bits/channel use at $14.5$ cm and $1.28$ bits/channel use at $25$ cm. Increasing the correction order to $N=35$ raises the lower bound to about $2.16$ bits/channel use at $\mu_x=0$ cm and $1.55$ bits/channel use at $25$ cm, but the persistent downward trend confirms that deterministic pointing bias remains a dominant impairment.
	
	\begin{figure}[!t]
		\centering
		\includegraphics[width=\columnwidth]{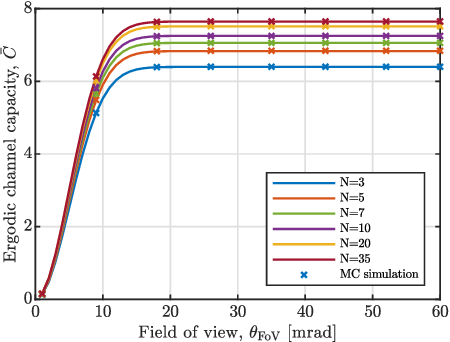}
		\caption{\ac{ealb} versus receiver \ac{fov} for the \ac{dl} case.}
		\label{fig:capacity_versus_field_of_view_downlink}
	\end{figure}
	
	\begin{figure}[!t]
		\centering
		\includegraphics[width=1.0\columnwidth] {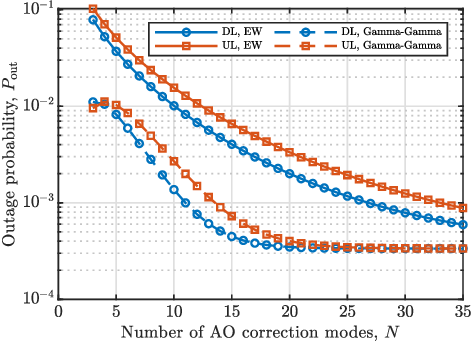}
		\caption{Outage performance comparison versus \ac{aoc} for current \ac{ew} model in this paper and \ac{gg} model given in Ref. \cite{ata2022haps}.}
		\label{fig:comparison_EW_GG}
	\end{figure}
	Fig.~\ref{fig:capacity_versus_field_of_view_downlink} shows the \ac{dl} \ac{ealb} versus receiver \ac{fov}. The lower bound is nearly zero at $\theta_{\mathrm{FoV}}\approx1$ mrad because little optical power is accepted, then rises rapidly; at $9$ mrad, it reaches about $2.55$, $2.75$, $2.85$, $2.9$, $3.0$, and $3.05$ bits/channel use for $N=3$, $5$, $7$, $10$, $20$, and $35$, respectively. Beyond $\theta_{\mathrm{FoV}}\approx18$ mrad, the curves saturate, meaning that the \ac{fov} limitation is removed and residual turbulence/link attenuation dominate. In this region, the lower bound increases from about $3.2$ bits/channel use for $N=3$ to $3.8$ bits/channel use for $N=35$.
	
	\figurename~\ref{fig:comparison_EW_GG} compares the proposed \ac{ao}-aware \ac{ew} model with the \ac{gg}-based \ac{haps}-\ac{fso} model in Ref. \cite{ata2022haps}. Both models improve with increasing \(N\), but the \ac{gg} curves approach the \ac{aoa}/\ac{fov} blocking floor more rapidly, while the \ac{ew} model decreases more gradually and remains more sensitive to the residual \ac{ao}-corrected turbulence. This is consistent with recent observations that \ac{ao}-corrected irradiance statistics may deviate from conventional \ac{gg} fading and can be better represented by the \ac{ew} distribution \cite{stotts2024PDFModels}.
	
	\begin{figure}[!t]
		\centering
		\includegraphics[width=\columnwidth]{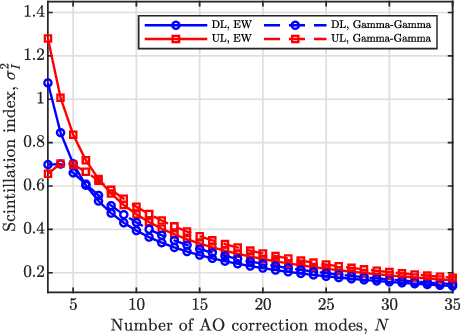}
		\caption{Comparison of \ac{ao}-dependent \acp{si} for the \ac{ew} and \ac{gg} turbulence models.}
		\label{fig:scintillation_comparison}
	\end{figure}
	
	Fig.~\ref{fig:scintillation_comparison} compares the \ac{ao}-dependent \ac{si} of the \ac{ew} and \ac{gg} models, where the solid and dashed curves denote $\sigma_{I,\mathrm{AO},\mathrm{EW}}^2(N)$ and $\sigma_{I,\mathrm{AO},\Gamma\Gamma}^2(N)$, respectively. Since $\sigma_I^2(D_G)$ is the tip/tilt reference at $N=3$, the decrease for $N>3$ confirms the scintillation-reduction effect of higher-order \ac{ao}. The \ac{gg} model gives slightly larger residual scintillation over most of the correction range, which partly explains its more conservative outage behavior.
	
	\section{Conclusions} \label{sec:conclusion}
	This paper investigated the performance of \ac{haps}-based \ac{fso} links with \ac{ao}-assisted turbulence mitigation. An \ac{ew}-based irradiance fading framework was adopted, where the effect of \ac{aoc} was incorporated through an effective residual scintillation parameter. Based on this model, analytical expressions were derived for the \ac{op}, average \ac{ber}, and \ac{ealb}, and the corresponding high-\ac{snr} asymptotic results were also provided. The analytical expressions were verified through \ac{mc} simulations generated under the same adopted composite channel model, showing close numerical agreement over the considered parameter ranges.
	
	The numerical results demonstrated that increasing the number of corrected \ac{ao} modes substantially improves the link performance for both \ac{ul} and \ac{dl} transmission. For example, increasing the correction order from \ac{ttc} to $N=35$ reduced the average \ac{ber} by nearly two orders of magnitude and improved the \ac{ealb} by more than $0.5$ bit/channel use in representative operating regimes. The outage analysis further showed significant \ac{snr} savings at a fixed \ac{op}, confirming the effectiveness of higher-order \ac{aoc} in suppressing deep fading events. The gains were most pronounced in moderate turbulence and moderate geometric-loss conditions, whereas performance saturation occurred at high correction orders or under severe slant-path propagation.
	
	The parametric studies also highlighted the key design tradeoffs for \ac{haps}-based \ac{fso} links. Larger zenith angles and higher \ac{haps} altitudes degraded the performance because of increased propagation distance and accumulated turbulence effects. Similarly, boresight error caused noticeable outage and achievable-rate degradation, showing that accurate pointing remains essential even when \ac{aoc} is employed. The beam-waist and \ac{fov} results further revealed that suitable optical beam shaping and receiver acceptance design are required to fully exploit the benefit of \ac{ao}. In particular, once the \ac{fov} becomes sufficiently large, the performance is mainly limited by residual turbulence and link attenuation rather than receiver angular acceptance.
	
	A comparison with the \ac{gg} reference model showed that the proposed \ac{ew}-based model provides a consistent and flexible description of the \ac{ao}-modified fading behavior, while the \ac{gg} approach generally leads to more conservative performance predictions. Overall, the results confirm that incorporating \ac{aoc} into the \ac{ew} fading model is an effective and analytically tractable approach for evaluating next-generation \ac{haps}-assisted \ac{fso} systems. Future work may extend this framework by using wave-optics-based \ac{ao} spatial filtering, adaptive mode selection, and experimentally measured turbulence data to further refine the residual scintillation modeling.
	

	\ifCLASSOPTIONcaptionsoff
	\newpage
	\fi

	\bibliographystyle{IEEEtran}
	\bibliography{biblio}

\end{document}